**Cell responses only partially shape cell-to-cell variations in protein abundances in *Escherichia coli* chemotaxis**

Sayak Mukherjee[1], Sang-Cheol Seok[1], Veronica J. Vieland[1,2,4], and Jayajit Das[1,2,3,5*]

[1]Battelle Center for Mathematical Medicine, The Research Institute at the Nationwide Children's Hospital and Departments of [2]Pediatrics, [3]Physics, [4]Statistics, [5]Biophysics Graduate Program, The Ohio State University, 700 Children's Drive, Columbus, OH 43205.

**Abstract:**

Cell-to-cell variations in protein abundance in clonal cell populations are ubiquitous in living systems. Since protein composition determines responses in individual cells, it stands to reason that the variations themselves are subject to selective pressures. But the functional role of these cell-to-cell differences is not well understood. One way to tackle questions regarding relationships between form and function is to perturb the form (e.g., change the protein abundances) and observe the resulting changes in some function. Here we take on the form-function relationship from the inverse perspective, asking instead what specific constraints on cell-to-cell variations in protein abundance are imposed by a given functional phenotype. We develop a maximum entropy (MaxEnt) based approach to posing questions of this type, and illustrate the method by application to the well characterized chemotactic response in *Escherichia coli* (*E. coli*). We find that full determination of observed cell-to-cell variations in protein abundances is not inherent in chemotaxis itself, but in fact appears to be jointly imposed by the chemotaxis program in conjunction with other factors, e.g., the protein synthesis machinery and/or additional non-chemotactic cell functions such as cell metabolism. These results illustrate the power of MaxEnt as a tool for the investigation of relationships between biological form and function.

**Significance Statement**

The relationship between form and function is ubiquitous in biology. Using a method (Maximum-Entropy) from statistical physics we investigated how function regulates form in the context of E. coli chemotaxis. We found that the nearly perfect and robust chemotaxis behavior (function) does not fully determine the cell-to-cell variations of chemotaxis protein abundances (form) in E. coli. We show that additional constraints imposed by the protein synthesis machinery and non-chemotactic cell functions in conjunction with the constraints imposed by the chemotaxis program are required to determine the observed variations of protein abundances. This demonstrates that properties of a modular component (e.g., the chemotaxis signaling module) in a biological network also depend on the system of which the module is a part.

Cell-to-cell variations in protein abundances or copy numbers are commonly found in genetically identical cells(1, 2). Because protein abundances directly regulate cell responses through signaling networks, a logical form-function relationship would imply that, in the context of an adaptive behavior of a cell population (function), these variations between individual cells (form) should themselves be under selection pressures. But functional implications of cell-to-cell variations of protein abundances are in general not well understood(3).

The relationship between form and function is an abiding theme of biological research(4, 5). The most common way to probe this relationship is through manipulation of form, e.g., perturbing parameters of the system and observing the effects on function (this can be done *in vivo, in vitro* or *in silico*). Experiments of this type in recent years have indeed demonstrated functional consequences of cell-to-cell variation in protein abundances: *e.g*., differences in protein abundances can produce distinct lineage commitments in haematopoietic stem cells(6); and covariation in protein abundances has been shown to increase the efficiency of chemotactic responses in *E. coli*(7, 8). This type of experiment can be illuminating but it can also be incomplete, because the range of perturbations considered is subject to practical limitations as well as the limits of our imaginations regarding what other possibilities exist.

Here we turn this procedure around and ask instead, if we start from an evolutionarily favored function, what general features of form must then exist? That is, rather than asking what is the impact on function of some selected features of form, we ask instead what constraints are imposed on form by selective factors operating at the level of function. In the context of cell-to-cell variations of protein abundances, this question becomes, how does the ability of individual cells to respond to changes in the local environmental shape the nature of variations of protein abundances in a cell population? Addressing this question is important to acquire a better understanding regarding the functional role of the cellular heterogeneity.

We develop a general methodology for asking such questions, based on MaxEnt(9, 10). To illustrate, we apply the method to one particular feature of form – cell-to-cell variation in protein abundances or copy numbers in genetically identical cells – in the context of a very well characterized, highly robust system - chemotaxis in *Escherichia* *coli* (*E. coli*). Using experimental data from cell population based assays as well as single cell experiments available in the published literature(11, 12) we use MaxEnt to ask two distinct, but related, questions. First, we investigate whether the observed chemotactic responses are sufficient to explain the variations in protein abundances. We find that the answer to this question is no, and hypothesize that additional constraints on the relationships among protein abundances are required, above and beyond the constraints inherent in the chemotactic response itself. The second question we ask relates to the nature of these additional constraints. We show that when constraints jointly imposed by the chemotaxis program itself and by cell functions and processes not directly related to chemotaxis (such as processes involved in protein synthesis) determine mean values and pair-correlations of the chemotaxis protein abundances in individual cells, the cell population remarkably reproduces the measured chemotactic responses. These results demonstrate the role of non-chemotactic functions in shaping the form of the chemotaxis signaling network module in E. coli, which is widely believed to be relatively isolated (13). This adds another important example of an emerging theme in biology, namely, that properties of a modular component in a biological network depend on the system of which the module is a part (14). In addition to shedding additional light on *E. coli* chemotaxis, these results illustrate the power of the MaxEnt methodology as a general tool for the investigation of relationships between biological form and function.

## Results

### *E. coli* chemotaxis and the MaxEnt method

Single E. coli cells sense the presence of attractants such as amino acids in the medium and swim toward the nutrient source. Upon reaching the region of higher nutrient concentrations the cells return to their pre-stimulus state of random movements displaying a nearly perfect adaptive behavior. In individual E. coli cells, membrane bound Tar receptors bind to attractant molecules and initiate a series of biochemical signaling reactions (Fig. 1A) leading to a transient dephosphorylation of the phosphorylated form of a key cytosolic protein CheY or CheY-P. CheY-P controls the direction of rotation (clockwise or anti-clock wise); a decrease in CheY-P abundance favors anti-clock wise rotations and propels the single cell toward the attractants (Fig. 1B). The CheY-P abundance slowly increases back to its pre-stimulus level as a the copy number of methylated Tar receptors, which lead to phosphorylation of CheY at an increased rate, is gradually elevated due to methylation reactions carried out by CheR in the presence of attractants (Fig. 1A). Throughout the signaling the methylation/de-methylation processes are executed by the enzymes CheR/CheB and the phosphorylation/de-phosphorylation processes are carried out by the enzymes CheA-P/CheZ. Since CheY-P abundances in single cells regulate flagellar rotations, the chemotactic response in single E. coli cells can be characterized by variables describing the time scale and the adaptive behavior of the CheY-P kinetics (Fig. 1B); specifically, (1) adaptation time, $\tau_\alpha$, defined as the time the CheY-P abundance in an E. coli cell (indexed by α) takes to rise up to half of its pre-stimulus value from the time when attractants were added, (2) precision of adaptation, $s_\alpha$, calculated as the absolute value of the relative difference in the steady state abundance of CheY-P (denoted by [CheY-P]) in a single E. coli cell at the pre ($[\text{CheY-P}]_{\text{pre-stim}}$) and post stimulation ($[\text{CheY-P}]_{\text{post-stim}}$ )conditions, i.e., $s_\alpha = |([\text{CheY-P}]_{\text{post-stim}} - [\text{CheY-P}]_{\text{pre-stim}})|/[\text{CheY-P}]_{\text{pre-stim}}|$. We also consider (3) the variation of the pre-stimulus steady state abundance of CheY-P (or $p_\alpha$) in an E. coli cell relative to its value at the optimal condition as a relevant variable for characterizing chemotactic responses for the reasons below. Previous experiments and mathematical models pioneered by Barkai and Leibler (15)(or the BL model from now on) (*SI Appendix*, Fig. S1, Tables S1-S2) demonstrated the robustness of the perfect adaptive nature of chemotactic response to variations of protein abundances and kinetic rates. However, single cell experiments of the E. coli flagellar motor response showed that the motor can work properly within a 30% variation from the optimal steady sate concentration of CheY-P at about 3μM(16). The BL model produces substantial changes to steady state CheY-P abundances against large variations of protein abundances and thus is unable to explain the robustness of the motor function for such large perturbations. Sourjik and colleagues(7) modified the BL model (the modified BL model or the MBL from now on) to account for the proper functioning of the flagellar motor. A key extension of the MBL model over the BL model was the inclusion of a CheZ dependent de-activation of CheY-P dephosphorylation (Fig. 1A).

We develop a MaxEnt based method to quantify the minimally structured cell-to-cell variations in total protein abundances required to reproduce the observed chemotactic responses in single cell and cell population based experiments. We considered cell-to-cell variations of total abundances of chemotaxis proteins (17, 18) as well as intrinsic fluctuations in copy numbers of signaling proteins within individual *E. coli* cells that arise due to the stochastic nature of biochemical chemotaxis signaling reactions(17-19). Upon addition of attractants in the medium at time $t = t_0$, in an individual cell containing total protein abundances given by $\{n^{\text{total}}{}_q\}$ ($q$=1… $N_T$, representing the chemotaxis proteins: Tar, CheA, CheB, CheR, CheZ, and CheY), the copy numbers of signaling molecules change with time due to the signaling reactions. We define a stochastic trajectory, Γ, representing changes in the abundances of signaling proteins with time in an individual cell by a set ($\{n_j\}, t_n ; \{n_j\}, t_{n-1} ; \{n_j\}, t_{n-2}; \ldots; \{n_j\}, t_1 ; \{n_j\}, t_0 ; \{n^{\text{total}}{}_q\}$) where copy numbers of different proteins, $\{n_j\}$ (j=1…$N_P$; $N_P$= total # of distinct signaling proteins) change at the times $\{t_{n-i} = t_0+(n-i)\Delta$, i=0..n}. Δ is taken to be smaller than or of the same order of the smallest reaction time scale (Fig. 2). $N_P \geq N_T$, as a protein species can be modified during signaling, e.g., CheY-P is generated from the protein CheY during signaling. We use the MaxEnt technique to estimate the probability distribution of these trajectories ($P_\Gamma$), specifically, by maximizing Shannon's entropy ($S$)

$$S = -\sum_{\Gamma} P_{\Gamma} \ln P_{\Gamma} \qquad (1)$$

in the presence of constraints imposed by experimental measurements pertaining to chemotactic responses or chemotaxis protein abundances. Eq. 1 is also known as the path entropy and the constrained maximum is also referred to as maximum caliber (MaxCal)(10). We carried out maximization of S in the presence of two types of constraints that capture relevant information(20) regarding E.coli chemotactic responses and the nature of the cell-to-cell variations of total protein abundances: (i) Constraints characterizing chemotactic responses. Since the essential features of chemotactic responses in a single cell (indexed by Γ) are described by the variables, $\tau_\alpha$, $s_\alpha$, and, $p_\alpha$, we used average values and variances of these variables over a cell population as constraints. (ii) Constraints describing the shape of cell-to-cell variations of total protein abundances. We used the average values as well as the variances and co-variances of the protein abundances as constraints. Since $P_\Gamma$ represents the joint distribution $P(\{n_j\}, t_n ;\{n_j\}, t_{n-1} ; \{n_j\}, t_{n-2}; \ldots; \{n_j\}, t_1 ; \{n_j\}, t_0 ; \{n^{\text{total}}{}_q\})$, any change in the shape of cell-to-cell variations of total protein abundances or $P(\{n^{\text{total}}{}_q\})$ will produce changes in $P_\Gamma$. We sought to estimate the maximally varying, or the least structured distribution $\hat{P}\left(\left\{n_q^{\text{total}}\right\}\right)$ consistent with constraints imposed by the available experimental data as described in (i) or (ii). (See Methods for explicit derivation of the underlying equations.) Therefore $\hat{P}\left(\left\{n_q^{\text{total}}\right\}\right)$ represents a probability distribution that is sufficient to characterize what is known about the underlying system (the constraints), without the imposition of any additional assumptions not directly justified by the available empirical data(9, 10). The constraints in (i) estimate the minimal structure imposed by the chemotactic responses themselves on the distribution of total protein abundances in E. coli cells, while the constraints in (ii) probe the minimal structure in cell-to-cell variations of total protein abundances that is able to reproduce the measured chemotactic responses while remaining consistent with the observed protein abundances. If the distribution of the protein abundances is entirely shaped by the chemotactic responses, then the estimated $\hat{P}(\{n_q^{\text{total}}\})$ will be the same using the constraints in (i) and the constraints in (ii). Our results show, however, that this is not the case. We describe our results in the next sections.

## Chemotaxis itself is not sufficient to explain observed protein abundance distributions

We maximized the entropy (Eqn. (1)) to evaluate (details in Methods section) the least structured cell-to-cell variations in protein abundances required to produce experimentally observed chemotactic response and then compared the inferred distribution with the available data pertaining to cell-to-cell variations of E. coli proteins from experimental observations. Specifically we compared the mean values, variances, and co-variances of the least structured distribution with the available measurements. Since all the six chemotactic proteins (Tar, CheA, CheR, CheB, CheZ, CheY) regulate the variables, $\tau_\alpha$, $s_\alpha$, and, $p_\alpha$, in single cells, constraining averages and variances of these variables, in principle, could in turn constrain variations of total protein abundances of the chemotactic proteins. When average values of $\{\tau_\alpha\}$(or $\overline{\tau}$), $\{s_\alpha\}$(or $\overline{s}$), and, $\{p_\alpha\}$(or $\overline{p}$) were constrained individually, the corresponding least structured distribution of protein abundances produced small correlations, both positive and negative, between the protein abundances (*SI Appendix*, Fig. S2). This was in stark contrast with the experimental observation that shows strong positive correlations (≈1) between the proteins CheY and CheZ or CheA and CheY (7, 8), or, between CheY and CheB as observed in vitro experiments using cloned gene pairs (8). The average values of the protein abundances in the inferred distribution showed much larger values compared to their experimental counterparts (*SI Appendix*, Table S3). When $\overline{\tau}$, $\overline{s}$, and, $\overline{p}$, were constrained at the same time, the qualitative features of the least structured distribution of protein abundances did not change (*SI Appendix*, Fig. S2). Including the variances of the variables, $\overline{\tau^2}$, $\overline{p^2}$, and $\overline{s^2}$ in the set of constraints in different combinations increased the magnitude of the correlations between the protein abundances, however, the correlations contained both positive and negative values and the average values of the protein abundances were still larger compared to their experimental counterparts (*SI Appendix*, Fig. S2-S3,Table S3).

We describe results from a particular case where the variables, $\overline{\tau}$, $\overline{\tau^2}$, $\overline{s}$, $\overline{p}$, and $\overline{p^2}$ were constrained in detail below to discuss specific agreements and disagreements of the variations of protein abundances with experimental observations and their biological implications. The above constraints produced correlated variations in the protein abundances (Fig. 2 and *SI Appendix*, Fig. S4), however, the magnitude of the correlations were smaller compared to that observed in experiments. Positive correlations were obtained between abundances of multiple protein pairs (Fig. 2 and *SI Appendix*, Table S4), including the protein pairs (e.g., CheY-CheZ, CheY-CheB, CheY-CheR, CheR-CheZ, Tar-CheR, Tar-CheB, Tar-CheZ) which are encoded by genes (*cheY*, *cheZ*, *cheB*, *mcp*, and *cheR*) residing in the *meche* operon(7, 8), and, the protein pairs (e.g., CheA-CheB, CheA-CheZ) that are encoded by genes residing in two different operons, *meche* and *mocha* (contains *cheA*)(7, 8). When *E. coli* chemotaxis proteins are encoded by genes in the same operon, they are translated by the same polycistronic mRNA(21), thus abundances of those proteins are likely to be positively correlated. Therefore, the observed positive correlations for the MBL model between the protein abundances encoded by the *meche* operon are qualitatively consistent with the notion of co-regulated gene expressions for the genes in the same operon. The positive correlation between CheY in CheZ is in direct qualitative agreement with single cell experiments measuring

CheY and CheZ simultaneously(7, 8). Positive MaxEnt correlations between other pairs of protein abundances have not been directly measured in single cell experiments. However, the predicted positive correlations between the abundances of CheY-CheB and CheY-CheZ are consistent with *in vitro* experiments assaying correlations in protein expressions using cloned gene pairs (Fig. 2 and *SI Appendix*, Table S4) (8). Most of the protein pairs producing positive correlations (*SI Appendix*, Table S4) in the inferred distribution also showed strong pair-wise co-occurrence of the encoding genes in 527 bacterial genomes containing at least one chemotaxis gene(8) supporting that the chemotactic functions partially produces the observed correlated variations between these protein pairs qualitatively.

But the MaxEnt model also produced *negative* correlations between protein abundances (Fig. 2) for protein pairs (CheB-CheZ, CheR-CheB, Tar-CheY) encoded by the *meche* operon and protein pairs (Tar-CheA, CheA-CheY, CheA-CheR) encoded by the *mocha* and *meche* operons. This disagrees with experiments (7), which have demonstrated positive correlation between abundances of CheA and CheY, and between CheR and CheB, and also disagrees with *in vitro* protein expression measurements for cloned gene-pairs (8). In addition, the negative correlations between CheB-CheZ, CheR-CheB, and Tar-CheY would seem to contradict the idea that genes in the same operon are likely to produce positive correlation between corresponding protein expressions.

Furthermore, by comparison with data from single cell experiments, the univariate MaxEnt distributions of protein abundances showed larger means, variances, and higher order moments for the abundances of CheY, CheZ, and CheA (Figs. 3A-3B, and *SI Appendix*, Figs. S5-S6). Similarly, the second and higher order moments calculated from the predicted joint distributions of CheY and CheZ (Fig. 3C) or CheA and CheZ abundances showed a similar increased spread of protein abundances in single cells compared to the experiments. In addition, the mean abundances for the other proteins in the inferred distribution were consistently larger than their measured counterparts. Constraining all the six variables, $\overline{\tau}$ , $\overline{\tau^2}$ , $\overline{p}$ , $\overline{p^2}$ , and $\overline{s}$ , $\overline{s^2}$ did not produce any qualitative change in the results (*SI Appendix*, Fig. S2 and Table S3). This shows that regardless of the combination of the constraints involving the variables describing the chemotactic responses consistently produces a broader distribution with larger mean values and positive and negative covariances and including additional constraints does not lead to a qualitatively better agreement between the inferred distribution and the experiments. This tells us that the distribution of protein abundances in E. coli is subject to *additional* constraints, not yet incorporated into the MaxEnt calculation.

Since the chemotactic program itself does not sufficiently constraint the protein abundance distribution, we hypothesize the importance of additional constraints arising from physical and biochemical processes that control synthesis, and other non-chemotactic functions of these different proteins in a cell. Taking clue from the result that a multivariate normal distribution can be used to reasonably approximate the inferred distribution $\hat{P}(\{n_q^{\text{total}}\})$ in protein abundances (Fig. 3D), we hypothesized that the efficient chemotactic program in individual E. coli cells along with processes not directly related to the chemotaxis regulate the mean values and the pair-correlations in the chemotaxis protein abundances. We turn to these in the next section.

**Mean values and pair-wise correlations in protein abundances regulate chemotactic responses in *E. coli*.**

Here we again use the MBL model, but we do not impose constraints on the chemotactic parameters. Instead, we introduce constraints directly on the protein abundances and compare the resulting MaxEnt model to the observed chemotactic response.

We first considered the MaxEnt distribution subject only to constraints on the means of the protein abundances taken from cell population measurements. The model showed exponentially distributed protein abundances with vanishing co-variations (details in the web supplement). This distribution generated chemotactic responses with values for $s$, $p$, and $\tau$ (in individual cells) which were substantially different than that observed in experiments for wild type *E. coli* (*SI Appendix*, Fig. S7, Table S5). We then further constrained the variances and co-variances between different protein abundances. Magnitudes of variances and co-variances for most of the chemotaxis protein pairs, except CheY-CheA and CheY-CheZ (7), were not directly available from the published experiments. However, as suggested in Ref.(7), a log-normal distribution for all proteins similar to that of CheY and CheZ reproduced the observed average values of s, p, and $\tau$, as well as their distribution reasonably well (*SI Appendix*, Fig. S7). Thus, we used covariances calculated from the log-normal distribution for those proteins abundances that have not been directly measured in single cell experiments.

With these additional constraints in place, the MaxEnt distribution was a multivariate normal for the protein abundances, and the resulting chemotactic response produced distributions of $s$, $p$, and $\tau$, that showed excellent agreement with experiments (Fig. 4A-4C, *SI Appendix*, Fig. S8). We tested this conclusion further by leaving covariances between different pairs of protein unconstrained. Our results show as long as covariances that minimally connect all the proteins abundances are constrained the generated chemotactic response is in reasonable agreement with the experiments (*SI Appendix*, Figs. S7 and Figs. S9-S10). This represents the minimal set of constraints in protein abundances required to produce the observed chemotactic response (see the supplementary material for details). This supports the hypothesis that a combination of constraints imposed by an efficient chemotactic response and other factors, e.g., the protein synthesis machinery and/or the non-chemotactic functions of the chemotaxis proteins determine the mean values, variances, and co-variances that minimally connect all the chemotaxis proteins abundances in the wild type E. coli.

**Discussion**

Examples of form constraining function are ubiquitous in living systems(5). Here we addressed the inverse question "how function constrains form" in the context of E. coli chemotaxis. Specifically, using a MaxEnt based approach we studied the minimal restriction imposed on cell-to-cell variations of protein abundances by the measured chemotactic response in individual E. coli cells. We found that the observed chemotactic response imposed both positive and negative co-variations between protein abundances (Fig. 2 and *SI Appendix*, Table S4). The positive correlations suggest that the requirement to execute an efficient chemotaxis program, crucial for increased growth or fitness of an *E. coli* population, leads to selection of processes that can co-regulate protein expressions in E. coli. Formation of operons could provide a potential mechanism to generate positively correlated protein expressions(21) as genes in an operon transcribed by the same mRNA are likely to produce co-regulated gene expressions. *E. coli* chemotaxis proteins are encoded by genes residing in two operons, *meche* and *mocha*(7, 8). Therefore, the positive co-variations of protein abundances in the MBL model in the pairs, CheY-CheZ, CheY-CheB, CheY-CheR, CheR-CheZ, Tar-CheR, Tar-CheB, and, Tar-CheZ, indicate that the requirement of efficient chemotaxis helped in formation of the *mocha* operon. Also, proteins encoded by genes in different operons can become correlated during translation (8, 22, 23). Therefore, the positive correlations between CheA-CheB or CheA-CheZ could assist in evolutionary selection of such processes.

An intriguing aspect of our results is the imposition of negative correlations in protein abundances by the observed chemotactic behavior. Functional implications of the negative correlations in the chemotaxis signaling kinetics is evident in most of the cases (*SI Appendix*, Table S4), e.g., the negative correlation between CheA and CheY abundances; since CheA activates CheY, an increase in the abundance of CheA accompanied by a decrease of CheY abundance keeps the abundance of CheY-P unchanged and thus increases robustness. Surprisingly, this result contradicted with single cell experiments that showed a positive correlation between the protein abundances(7). Furthermore, the same minimally structured distribution produced much larger mean values and higher order moments compared to the experiments (Fig. 3). This led us to hypothesize that mean values and the pair-correlations in protein abundances whose primary function is to execute chemotactic signaling in E. coli are largely determined by constraints imposed both by an efficient chemotaxis program and functions directly unrelated to chemotaxis (Fig. 5). Limitations such as, finite pool of RNA polymerases and ribosomes in individual cells(24, 25) or energetic costs for protein synthesis(26), imposed by the protein synthesis machinery could restrict protein expressions. These restrictions are manifested in reduction in cell growth rate when abundances of non-functional proteins are increased in bacterial cells(24, 27). A tight regulation of protein abundances is also demonstrated in the results showing that the steady state protein abundances are remarkably conserved across species(28). Moreover, a chemotaxis protein could be involved in non-chemotactic function as well such as cell metabolism, e.g., CheY is linked with the metabolic state of the cell(29).

We found that the mean values and the pair-correlation functions between protein abundances that minimally connect all the protein abundances are required to be constrained to be able to produce the measured chemotactic responses in individual *E. coli* cells (*SI Appendix*, Figs. S7-S8). In this case, all the pair correlations needed to be constrained to large positive values. Therefore, these non-chemotactic functions lead to selection of a distribution of protein abundances that links all the chemotactic proteins simultaneously with strong positive correlations. This is chosen over the distribution containing weaker pair-correlations with positive and negative values as preferred by the chemotactic responses alone. This result, in addition to emphasizing the role of non-chemotactic functions in shaping protein abundances involved in a relatively insulated chemotaxis signaling module, also lends support to an emerging theme in biology that properties of a biological module can be influenced by the system where the module is embedded in(14).

Recent work by Salman et al. (30) showed that protein abundances in E. coli and yeast that are involved in metabolism can be scaled to a 'universal' non-Gaussian scaling function when protein abundances are scaled with the mean values and the variances. This result urges us to speculate if the scaling of distributions of protein concentrations to a 'universal' scaling function reflects the adequacy of the mean values and the pair-wise correlations to produce the essential variations in the phenotype primarily regulated by those proteins in individual cells. It will be worthwhile to investigate the generality of these results for other phenotypes in other cell types. The proposed MaxEnt method is general and can be employed to probe such function-form relations in other living systems.

**Materials and Methods:**

**Calculation of MaxEnt (MaxCal) solutions:** We seek to determine that least structured distribution of total protein abundances or $\hat{P}(\{n_q^{\text{total}}\})$ that maximizes $S$ in Eq. (1) in the presence of constraints imposed by the chemotactic response in E. coli. $P_\Gamma$ is related to $P(\{n^{\text{total}}{}_q\})$ by the relation,

$P_\Gamma = P(\{n_j\}, t_n ; \{n_j\}, t_{n-1} ; \ \dots ; \{n_j\}, t_1, \{n_j\}, t_0 \,|\, \{n^{\text{total}}{}_q\})\, P(\{n^{\text{total}}{}_q\}) = P_C\, P(\{n^{\text{total}}{}_q\})$ (2)

where, $P_C = P(\{n_j\}, t_n ; \{n_j\}, t_{n-1} ; \ \dots ; \{n_j\}, t_1, \{n_j\}, t_0 \,|\, \{n^{\text{total}}{}_q\})$ is the conditional probability of occurrence of the trajectory, $\Gamma_C$, represented by the set, $(\{n_j\}, t_n ; \{n_j\}, t_{n-1} ; \dots ; \{n_j\}, t_1, \{n_j\}, t_0)$ for a specific choice of total protein abundances, $\{n^{\text{total}}{}_q\}$. When a variable, $f_\Gamma$, describes a chemotactic response (such as $\tau$) that depends on the stochastic trajectory Γ produced in a single E. coli cell indexed by α, the cell population averaged value of $f_\Gamma$ is given by:

$$\frac{1}{\text{total \# of cells}} \sum_{\alpha=1}^{\text{total \# of cells}} f_\alpha = \sum_{\Gamma} f_\Gamma P_\Gamma = \overline{f}^{\text{expt}} \tag{3}$$

where $\overline{f}^{\text{expt}}$ denotes the average value of $f$ measured in experiments. We show the result that $P(\{n_q^{\text{total}}\})$ maximizes S in Eq. (1) for the constraint in Eq. (3) for simplicity. The result including additional constraints is shown in the supplementary material.

Since $P_\Gamma$ depends on $P(\{n^{\text{total}}_q\})$ via Eq. (2), it is possible to choose different shapes of $P(\{n^{\text{total}}_q\})$ which will satisfy the constraints imposed by Eq.(3). We seek to estimate the maximally varying or the least structured distribution $P(\{n^{\text{total}}_q\})$ where the minimal structure in the distribution arises solely due to the constraints imposed. For the constraint in Eq. (3), the $P(\{n^{\text{total}}_q\})$ that maximizes S (Eq. (1)) is given by (for a detailed derivation and discussion see the supplementary material),

$$P(\{n_q^{total}\}) = Z^{-1} Q_C \exp\left[-\lambda \sum_{\Gamma_C} f_\Gamma P_C\right] \tag{4}$$

where $\ln Q_C = -\sum_{\Gamma_C} P_C \ln P_C$. The sum over $\Gamma_C$ essentially denotes averages over variations of stochastic trajectories due to intrinsic noise fluctuations. The conditional probability $P_C$ can be calculated by solving the Master Equation(19) describing the biochemical reactions in the signaling model. $Q_C$ is then calculated from $P_C$. The Lagrange multiplier λ is calculated by substituting the estimated $\hat{P}(\{n_q^{\text{total}}\})$ in the constraint equation (Eq. (3)) and then solving the resulting nonlinear equation. We could also extend this method to a more general scenario, where the underlying intrinsic fluctuations are not quantifiable due to uncharacterized interactions in the signaling network. In such cases, $P_C$ can be inferred by imposing further constraints on $\ln Q_C$, provided data from repeated experiments on the same sample (or individual cell)(34) are available. For additional details see supplemental material (section IIIA) and Refs. (31-33).

When the mean values and the higher order moments of the total protein abundances are constrained instead of the chemotactic responses, the minimally structured distribution $\hat{P}\left(\left\{n_q^{\text{total}}\right\}\right)$ is calculated by maximizing the entropy

$$S^{\text{total}} = -\sum_{\{n_q^{\text{total}}\}} P(\{n_q^{\text{total}}\}) \ln(P(\{n_q^{\text{total}}\}) \tag{5}$$

instead of Eq. (1) as the structure of $P(\{n^{total}_q\})$ is independent of the chemotactic response in this case. The estimation of $\hat{P}\left(\left\{n_q^{total}\right\}\right)$ for this case is detailed in the supplementary material.

In our simulations we evaluate the distribution $\hat{P}(\{n_q^{total}\})$ in Eq. (4) in the following way. First we generate *a priori* distribution $Q(\{n_q^{total}\})$ by drawing total protein abundances from a uniform distribution $U(0, U_H)$, where $U_H$ is chosen to be roughly 10 times larger than the experimentally measured mean abundance of the corresponding chemotactic protein (35). Then the signaling kinetics in each individual cell is simulated by solving the deterministic biochemical reactions for the MBL model in the pre-stimulus condition (zero attractant concentration) using the rule based software package BIONETGEN(36). Once the kinetics reaches the steady state, attractants are added in the medium and the stochastic kinetics of the signaling reactions are simulated using BIONETGEN for a long time when the kinetics reaches a steady state. In the sample size (~70,000 single cells) we considered, each *E. coli* cell produces a unique chemotactic response composed of a stochastic trajectory Γ describing time evolution of abundances of signaling proteins, therefore, we identified each trajectory by the single cell that generated it (Fig. 1). The summation over $\Gamma_C$ in Eq. (4) is performed using this unique association of any trajectory with a single cell. Further details regarding the numerical scheme for constructing $\hat{P}(\{n_q^{total}\})$ in Eq. (4) are provided in the supplementary material (*SI Appendix*, sec III and Figs. S11-S13). We carry out simulations for the BL and the FT (37) model following the same scheme.

**Data from *E. coli* experiments:** The distribution of τ was obtained from Min et al. (38) by digitizing Fig.3C in that paper using an online web plot digitizer (http://arohatgi.info/WebPlotDigitizer/). The values of $\overline{\tau}$ and $\overline{\tau^2}$ are calculated from the distribution thus obtained. $\overline{s}^{expt}$ for wildtype RP437 strain was obtained from Alon et al (39). The average values of the chemotactic protein abundances were taken from Li et al. (35). The single cell distribution of CheY, CheZ, CheY-CheZ, and CheA-CheY for the wildtype RP437 strain were extracted from Kollman et al. (7) using the same graph digitizer.

**ACKNOWLEDGEMENTS**
This work was supported by funding from the Research Institute at Nationwide Children's Hospital and NIH grant AI090115 to J.D., and NIH grant MH086117 to V.J.V. J.D. thanks C. Jayaprakash for discussions. We thank the two anonymous reviewers for making constructive suggestions.

**Figure Captions**

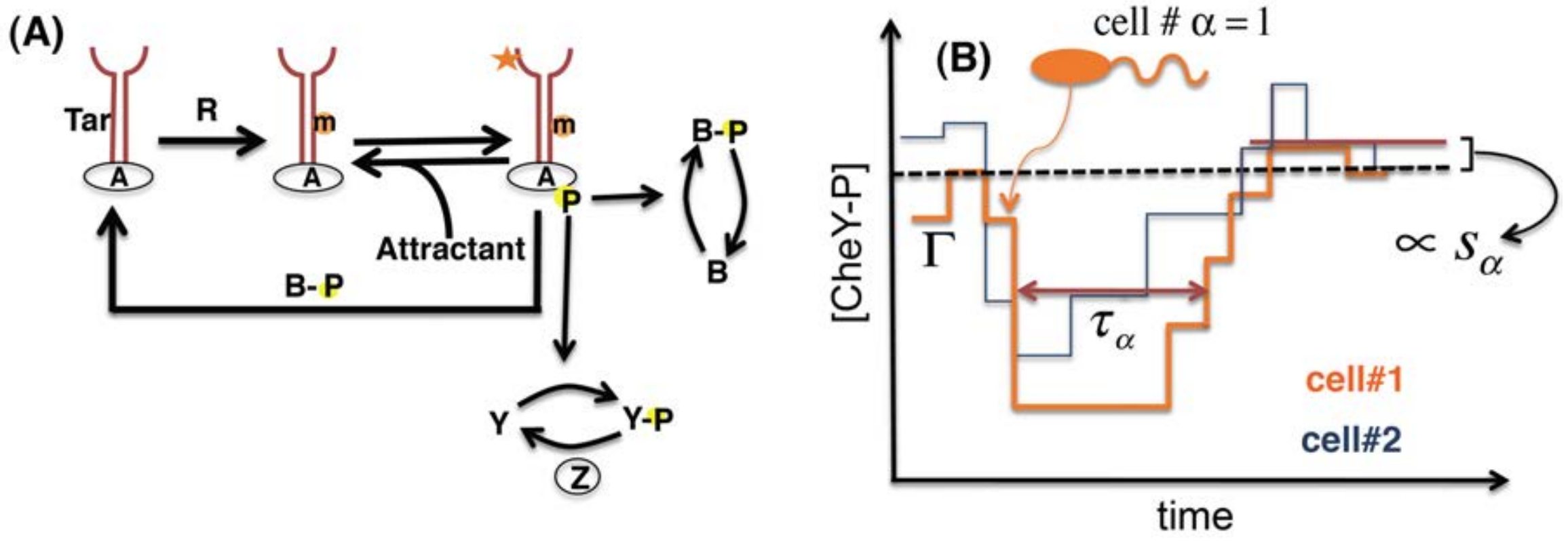

**Fig. 1. The chemotactic response in *E. coli***: (A) The chemotaxis signaling network for the MBL model. (B) Adaptive kinetics of copy number of CheY-P varies from cell to cell due to variations of protein abundances in individual cells as well as due to intrinsic noise fluctuations in the signaling reactions. The signaling proteins follow a unique stochastic trajectory, $\Gamma$, describing the kinetics of chemotaxis signaling in an individual E. coli cell indexed by α. The dashed line shows the cell population averaged value of steady state CheY-P abundance. For each stochastic trajectory, we calculate the adaptation time $\tau_\alpha$ (or $\tau_\Gamma$) the precision of adaptation, $\tau_\alpha$ (or $s_\Gamma$), and, the percentage variation of steady state abundance of CheY-P, $p_\alpha$ (or $p_\Gamma$) (not shown in the figure).

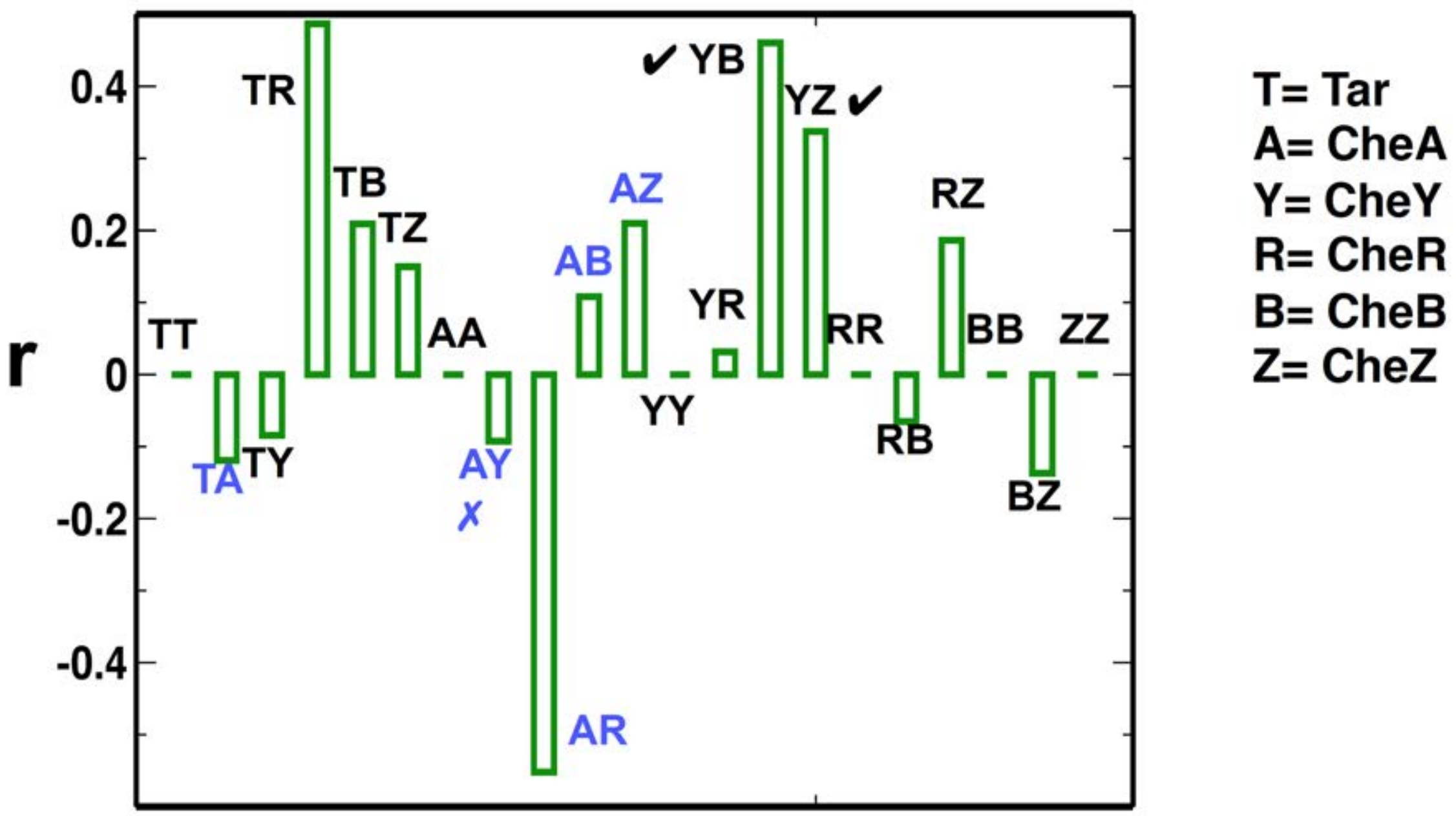

**Fig. 2. The observed chemotactic response imposes pair-wise correlations between proteins.** We constrained mean values of $\tau$, $\tau^2$, $s$, $p$, and, $p^2$ to the respective values measured in experiments, i.e., $\overline{\tau}$ = 245 s, $\overline{\tau^2}$ = 62323.5 $s^2$, $\overline{s}$ = 0.02, $\overline{p}$ =20%, and $\overline{p^2}$ = 425(%)$^2$, for estimating distributions of protein abundances using our MaxEnt approach. A) The pairwise Pearson correlation co-efficients are calculated for 6 chemotactic proteins using the MaxEnt distributions ($r_{\mathrm{MaxEnt}}$) for the MBL model and the *a priori* uniform distribution ($r_{\mathrm{uni}}$). We show the difference, $r = r_{MaxEnt} - r_{uni}$ for different protein pairs. $r_{\mathrm{uni}} \approx 0$ when cross correlations between protein pairs are considered. Since by definition, $r_{\mathrm{MaxEnt}} = r_{\mathrm{uni}} = 1$, r=0 when correlations between the same protein pairs (or variances) are considered. The protein pairs encoded by genes in the same and different operons are shown in black and blue, respectively. The agreement and disagreement with experiments assaying protein expression in single cells and in vitro cloned gene pairs are shown with a tick and a cross symbol, respectively.

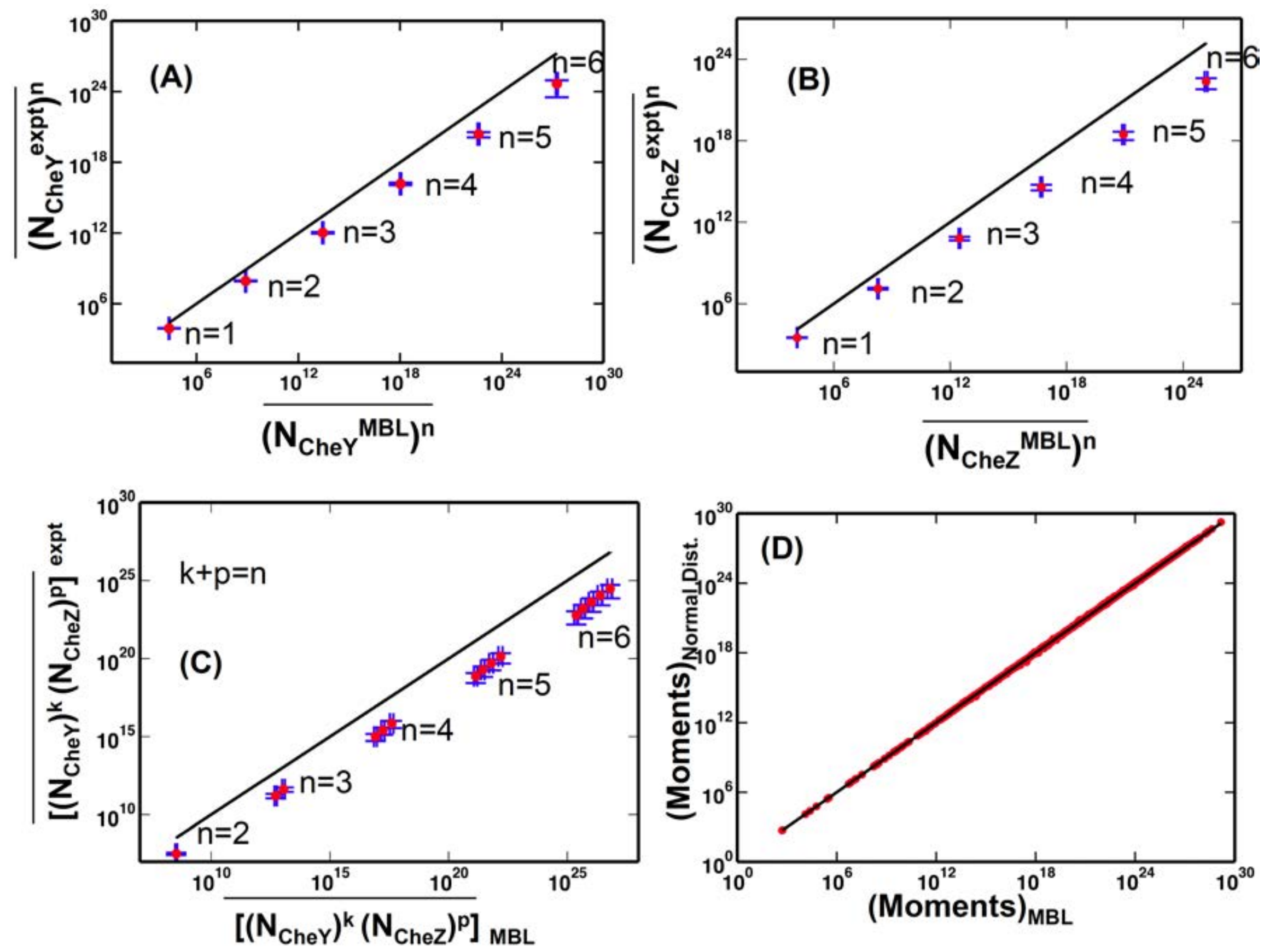

**Fig. 3. The minimally structured distribution of protein abundances enforced by E. coli chemotaxis is broader compared to the observed cell-to-cell variations.** (A) Comparison of the moments of CheY abundances calculated from the MaxEnt distribution with the data from single cell experiments in Ref. (7). The y=x line (solid black) is shown for comparison. (B) Similar comparison as in (A) for CheZ abundances. (C) Similar comparison as in (A) for the joint distribution of CheZ and CheY abundances. We further quantify the differences between inferred distribution and the experimental observations using $\chi^2$ (*SI Appendix*, Table S6) (D) Comparison of the MaxEnt distribution with a multivariate normal distribution. The multivariate distribution is constructed with the mean values and pair correlations equal to that of the MaxEnt distribution. We calculate all the moments up to the sixth order for all the six proteins for the MaxEnt and the constructed multivariate normal distribution. The y=x line (solid black) is shown for comparison.

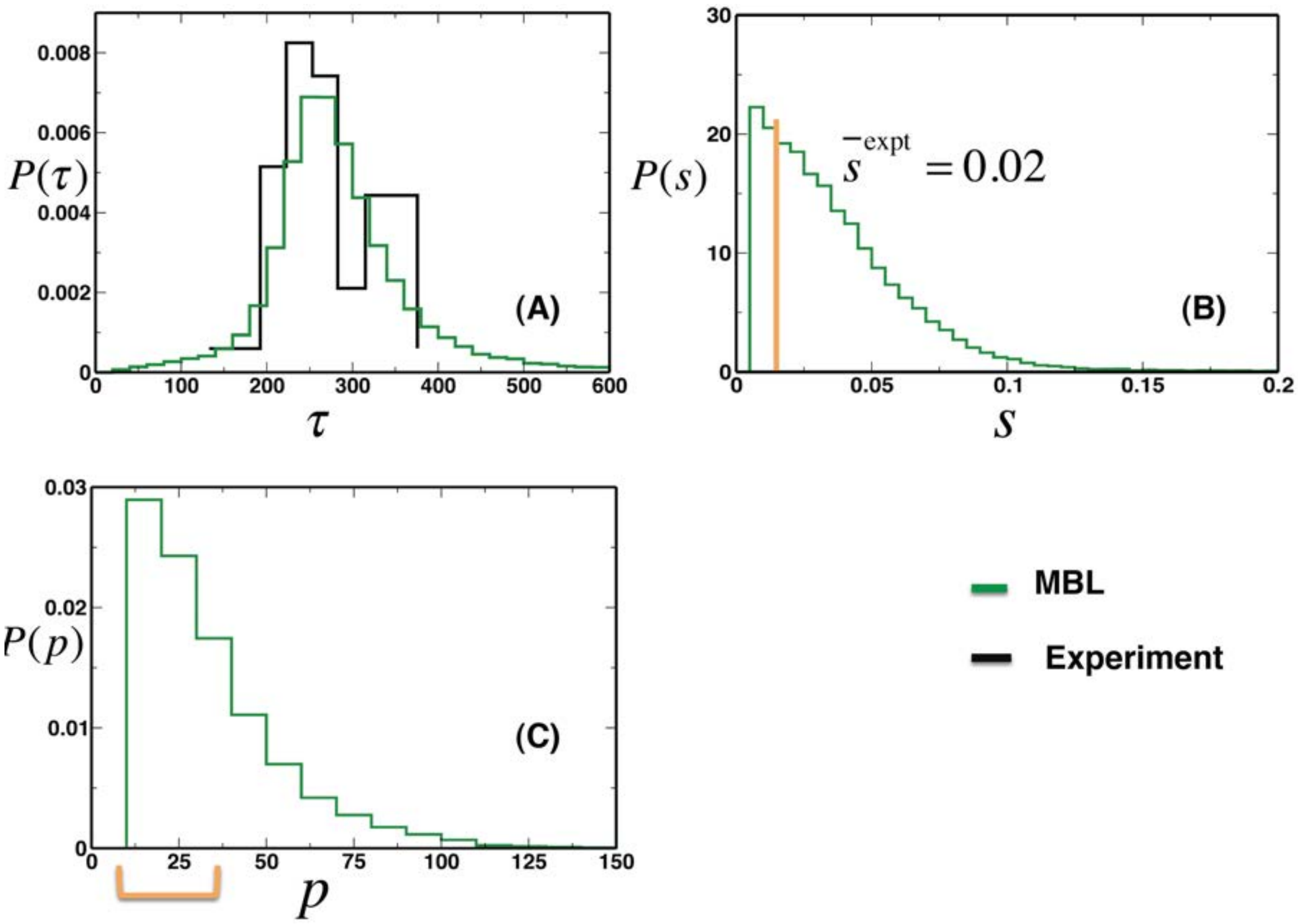

**Fig. 4. Pairwise correlations between protein abundances produce remarkable agreement between the predictions for chemotactic response and experiments in single cells.** (A) The distribution of the adaptation time for MBL is shown along with the experimental data (average value= 245 s, black stairs). (B) The distribution of the precision of adaptation, *s*, for the MBL model. The orange bar indicates the average precision of adaptation observed in wildtype RP437. (C) The distribution of the percentage variation *p* in the pre stimulus steady state of CheY-P abundance measured from an optimal value (see the Materials and Methods section and the supplementary material for details). The distribution shows that 70% of the cells are within the working range (p=30%) of the flagellar motor. The allowed range of percentage variation is shown with an orange bar.

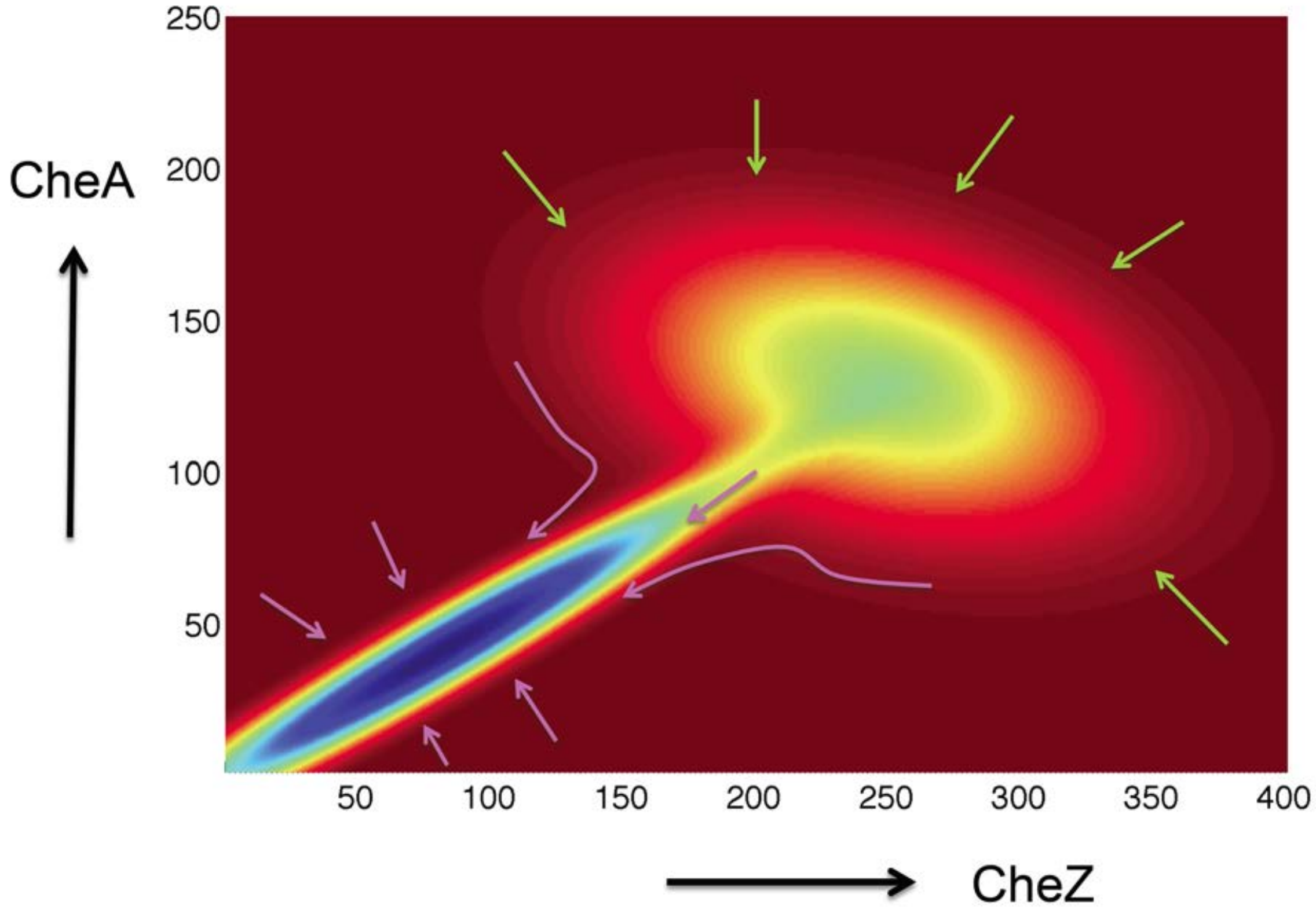

**Fig. 5. The chemotaxis program in combination with non-chemotactic phenotypes shape the cell-to-cell differences in chemotaxis protein abundances.** A schematic diagram shows the dependence of the E. coli fitness (z axis, cooler colors indicate higher values) landscape on variations of CheY and CheA abundances in individual cells. The observed chemotactic response leads (indicated by the green arrows) to selection of processes that impose correlations between the abundances. However, biochemical and biophysical processes regulating synthesis of chemotactic proteins and additional non-chemotactic cell functions where chemotaxis proteins also participate, drive (indicated by the pink arrows) the E. coli cells to a higher fitness value at lower and more restricted values of protein abundances.

**Supplementary Material for "Cell responses only partially shape cell-to-cell variations in protein abundances in *Escherichia coli* chemotaxis."**

**I. Models of E. coli chemotaxis:** We studied three different coarse-grained or approximate models that were proposed to explain E. coli chemotaxis: (i) the modified BL (MBL) model (1), which is a recently proposed modification of the BL model (2), (ii) the Barkai-Leibler (BL) model (2), and (iii) the fine-tuned (FT) model (3, 4). The biochemical reactions describing the models are shown in Fig. S1. Each model is composed of a set of biochemical reactions involving the chemotactic proteins, CheA, CheR, CheB, CheY, CheZ (not considered in the BL model), and the transmembrane Tar receptors. The models also differ from each other due to differences in their molecular wiring (Fig. S1). We provide details regarding each model below. The average abundances of the proteins in all the three models are shown in Table S1A. The activation probabilities for the receptors are shown in Table S1B.

(i) _MBL model:_ Sourjik and colleagues proposed and experimentally tested a modified version of the BL model (MBL, Fig. S1C) that was able to restrain variations in the steady state of CheY-P concentration to a small range (1). The biochemical reactions and the rates used are given in Table S2A.

_(ii) *FT model:*_ The FT model (Fig. S1A) was among the first proposed models that could explain precise adaptation in *E. coli* only within a narrow range of rate constants and protein concentrations (3, 4). We constructed the FT model by adding an extra module to the MBL model where the CheB-P can de-methylate the inactive forms of the Tar receptor complex ($Tar_m$ as opposed to $Tar_m^A$) as well. The reactions and the parameters used in the model are shown in Table S2B. Our version of the FT model, though very similar in essence to the original Knox et al. (4) and Hauri and Ross (3) model, harbors a few differences worth mentioning. Hauri and Ross approximated the Tar, CheW and CheA into one complex, which they referred to as T. They did not consider an explicit auto phosphorylation of CheA. We have an explicit auto phosphorylation reaction of CheA with a rate proportional to the total number of active Tar complex. Unlike Hauri and Ross model, CheA, instead of the complex T, transfers the phosphate group to CheY and CheB. We have explicitly considered the phosphatase CheZ. Hauri and Ross assumed a first order de-activation of CheY-P.

(iii) _BL model_: The BL model (Fig. S1B) was able to capture the robustness of the nearly perfect nature of adaptation of *E. coli* chemotaxis to variations in rate constants or strengths of protein-protein interactions (2), as confirmed in cell population experiments (5). However, the steady state concentration of phosphorylated CheY protein (or CheY-P) in the BL model is sensitive to large variations of protein concentrations in the model.

Since the working range of the flagella motor is limited to small variations (~30%) from the optimal CheY-P concentration, the robustness of *E. coli* adaptation to large range of variation in the abundance of chemotactic proteins requires small variations of steady state CheY-P concentrations(1, 6). The biochemical reactions and the rates used are given in Table S2C.

**II. In silico simulations of E. coli chemotaxis:** We used a rule based software package BIONETGEN(7) to simulate the deterministic and stochastic kinetics described by the biochemical reactions associated with each model. The signaling molecules are assumed to be well mixed in the simulation box representing a single E. coli cell.

*(A) Evaluation of the pre-stimulus state:* All the simulations were initialized at $t^{\text{initial}}$ = -800000 s with protein abundances, $\text{Tar}_{m=0}(t^{\text{initial}}) = \text{Tar}^{T}$, $\text{CheA}(t^{\text{initial}}) = \text{CheA}^{T}$, $\text{CheR}(t^{\text{initial}}) = \text{CheR}^{T}$, $\text{CheB}(t^{\text{initial}}) = \text{CheB}^{T}$, $\text{CheY}(t^{\text{initial}}) = \text{CheY}^{T}$ and $\text{CheZ}(t^{\text{initial}}) = \text{CheZ}^{T}$ while abundances of all other species (methylated receptors (m≠0) and phosphorylated form of all the other proteins) are set to zero. The superscript T refers to total abundances of the respective proteins in a single *E. coli* cell. We solve the ODEs describing the deterministic chemotactic kinetics to obtain steady state values of the protein abundances. The steady state values of the abundances define the pre-stimulus state of the chemotactic kinetics. The ODEs, instead of the stochastic kinetics, were used to make the calculations computationally efficient because the contributions of the intrinsic noise fluctuations to the steady state values of the ODEs were small (Supplementary Figure 4 in Mukherjee et al. (8)). The BIONETGEN files for the simulations are available at http://planetx.nationwidechildrens.org/~jayajit/Ecoli_chemotaxis_bionetgen_codes/.

*(B) Simulation of the chemotactic response:* The steady state protein abundances in the pre-stimulus state are used as initial conditions. The attractants (100μM of L-aspartate) are added in the system at $t = t_0 = 0$. We do not consider explicit ligand receptor interaction. The effect of ligand binding is subsumed in the probabilities a(L) of the methylated receptors to be in an active state (Table S1B). The stochastic simulations solve the Master Equation associated with the biochemical reactions exactly following the Gillespie method. The adaptation time ($\tau_\Gamma$), variation in the steady state CheY-P abundance ($p_\Gamma$), and the precision of adaptation ($s_\Gamma$) are calculated for each stochastic trajectory $\Gamma$ representing the chemotactic response in an individual cell. The stochastic simulations are carried out until $t$ =2000s which is much larger than the typical adaptation time for E.coli for a 100μM L-aspartate stimulation. When the CheY-P abundance in an individual cell does not recover to the half of the pre-stimulus steady state CheY-P abundance within 2000 s, we assign a very large number ($6 \times 10^{6}$ s) to $\tau$ to mark the cell that did not adapt in a realistic time scale. The pre-stimulus steady state abundance of CheY-P ([CheY-P]) varies from cell to cell due to the variations of total protein abundances in individual *E. coli* cells. The variation of the steady state abundance of CheY-P needs to be within 30% from an optimal value for proper functioning of the flagellar motor (1, 6). We calculate the variation of steady state CheY-P abundance ($p$) in single *E. coli* cells using the equation below, $\mathrm{p} = \left| \frac{[\mathrm{CheY-P}]_{\mathrm{pre-stim}} - [\mathrm{CheY-P}]_{\mathrm{optimal}}}{[\mathrm{CheY-P}]_{\mathrm{optimal}}} \right|$.

Optimal value of CheY-P ($[\mathrm{CheY\text{-}P}]_{\mathrm{optimal}}$ ), is defined as the ODE based solution of the steady state value of the CheY-P when the total protein concentrations are set to the values quoted in Li et al (9). We calculate $[\mathrm{CheY\text{-}P}]_{\mathrm{pre\text{-}stim}}$ using the ODE solutions ignoring intrinsic noise fluctuations for the reasons mentioned above.

**III. Evaluation of the distributions corresponding to the MaxEnt state.** MaxEnt provides a powerful tool to address such questions that require inference of model parameters based on available experimental data. The concept of the Maximum Entropy (MaxEnt) principle was seeded in the second law of thermodynamics which states that the entropy of an isolated system in equilibrium never decreases(10). The use of MaxEnt to infer distribution functions without imposition of any additional assumptions was pioneered by E. T. Jaynes and others (11, 12) where the Shannon's entropy (S) is maximized subject to constraints imposed by experimental data. These methods have been widely used in diverse disciplines including physics(10), information theory(13), econometrics, and, biology (14-17) to estimate probability distributions that are consistent with available experimental data (14, 16-18). We use an extension of the MaxEnt based approaches where Shannon's entropy constructed from distributions of temporal profiles of abundances of chemotactic signaling proteins is maximized. This method is known as the Maximum Caliber (MaxCal), also proposed by Jaynes to extend the MaxEnt approach to dynamical systems(12, 18, 19).

(A) Derivation of the inferred distribution $\hat{P}(\{n_q^{\mathrm{total}}\})$ when the chemotactic responses are constrained

We describe our method of inferring distributions of total abundances of protein species in individual E. coli cells below. We define the Shannon's entropy for the stochastic trajectories $\{\Gamma\}$ as,

$$S = -\sum_{\Gamma} P_{\Gamma} \ln P_{\Gamma} \qquad \text{(S1)}$$

When attractants are added at time $t=t_0$, the stochastic trajectory, $\Gamma$, represents changes in the abundances of signaling proteins in an individual cell in a time interval $t_0$ to $t_n$ by a set $(\{n_j\}, t_n ; \{n_j\}, t_{n-1} ; \{n_j\}, t_{n-2} ;\ldots; \{n_j\}, t_1;\{n_j\}, t_0; \{n^{\mathrm{total}}_q\})$ where copy numbers of different proteins, $\{n_j\}$ (j=1…$N_P$ = total # of distinct signaling proteins), are given at times, $t_{n-i} = t_0+(n-i)\Delta$, $i=0..n$, where, $\Delta$ is smaller than or of the same order as the smallest reaction time scale (Fig. 2). q denotes the number of different protein species, $N_T$. $N_P \geq N_T$, as a protein species can be modified during signaling, e.g., the signaling protein CheY-P is generated from the protein CheY.

Therefore,

$$\begin{aligned} P_\Gamma &= P(\{n_j\},t_n;\{n_j\},t_{n-1};\{n_j\},t_{n-2};\cdots|\{n_j\},t_0)P(\{n_j\},t_0;\{n_q^{total}\}) \\ &= P(\{n_j\},t_n;\{n_j\},t_{n-1};\{n_j\},t_{n-2};\cdots|\{n_j\},t_0)P(\{n_j\},t_0|\{n_q^{total}\})P(\{n_q^{total}\}) \end{aligned} \quad \text{(S2)}$$

$P(\{n_j\}, t_n ; \{n_j\}, t_{n-1} ; \{n_j\}, t_{n-2} ;\ldots; \{n_j\}, t_1 \mid \{n_j\}, t_0)$ is the conditional probability of producing the copy numbers of the signaling species in a stochastic trajectory Γ at the times $\{t_n \ldots t_1\}$, given there is a specific set of copy numbers of proteins ($\{n_j\}$) when attractants are added at the pre-stimulus state at $t_0$. $P(\{n_j\}, t_0; \{n^{total}_q\})$ denotes the joint probability of having the pre-stimulus state with specific copy numbers ($\{n_j\}$) at time $t_0$ and total protein concentrations $\{n^{total}_q\}$. This joint probability can be written as a product of the conditional probability $P(\{n_j\}, t_0 | \{n^{total}_q\})$, describing the probability for having the pre-stimulus state at $t_0$ given a specific set of total protein abundances $\{n^{total}_q\}$, and the probability of occurrence of $\{n^{total}_q\}$ or $P(\{n^{total}_q\})$, i.e, $P(\{n_j\}, t_0; \{n^{total}_q\}) = P(\{n_j\}, t_0 | \{n^{total}_q\})\, P(\{n^{total}_q\})$.

The biochemical signaling reactions producing E. coli chemotaxis are described by Markov processes where the conditional probability, $P(\{n_j\}, t_p | \{n_j\}, t_{p-1})$, for changing the signaling state of the system changes from $\{\{n_j\}, t_{p-1}\}$ to $\{\{n_j\}, t_p\}$ is given by the Master Equation(20),

$$\partial P(\{n_j\};t_p|\{n_j\};t_{p-1})/\partial t_p = L\,P(\{n_j\};t_p|\{n_j\};t_{p-1}) \quad \text{(S3)}$$

, where, $L$ describes a linear operator (20) dependent on the biochemical reaction rates, wiring of the signaling network in a signaling model (e.g., MBL), and the copy numbers of signaling proteins at time $t_{p-1}$. Therefore, the conditional probability, $P(\{n_j\}, t_n ; \{n_j\}, t_{n-1} ; \{n_j\}, t_{n-2} ;\ldots; \{n_j\}, t_1 \mid \{n_j\}, t_0) = P(\{n_j\}, t_n \mid \{n_j\}, t_{n-1})P(\{n_j\}, t_{n-1} \mid \{n_j\}, t_{n-2}) \ldots P(\{n_j\}, t_1 \mid \{n_j\}, t_0)$ (equality holds for a Markov process), is entirely determined by the solutions of the above Master Equation and the initial condition at $t=t_0$. We consider variations in $P_\Gamma$ arising from the variations in $P(\{n^{total}_q\})$, i.e.,

$$\delta P_\Gamma = P(\{n_j\},t_n;\{n_j\},t_{n-1};\{n_j\},t_{n-2};\cdots|\{n_j\},t_0)P(\{n_j\},t_0|\{n_q^{total}\})\delta P(\{n_q^{total}\}) = P_C\ \delta P(\{n_q^{total}\}) \quad \text{(S4)}$$

where, we define

$$P_C = P(\{n_j\},t_n;\{n_j\},t_{n-1};\{n_j\},t_{n-2};\cdots|\{n_j\},t_0)P(\{n_j\},t_0|\{n_q^{total}\})$$
$$= P(\{n_j\},t_n;\{n_j\},t_{n-1};\{n_j\},t_{n-2};\cdots;\{n_j\},t_0|\{n_q^{total}\})$$

to simplify notations.

We maximize S in Eq. (S1) when the total protein abundances are varied as above in the presence of the constraint given by Eq. (S5) below.

When a variable, $f_\Gamma$, describes a chemotactic response (such as τ) that depends on the stochastic trajectory Γ produced in a single E. coli cell indexed by α, the cell population averaged value of $f_\Gamma$ is given by:

$$\frac{1}{\text{total \# of cells}}\sum_{\alpha=1}^{\text{total \# of cells}} f_\alpha = \sum_\Gamma f_\Gamma P_\Gamma = \overline{f}^{\text{expt}} \tag{S5}$$

When $P_\Gamma = \hat{P}_\Gamma$ or $P(\{n_q^{total}\}) = \hat{P}(\{n_q^{total}\})$ maximizes the entropy S in Eq. (S1), small variations of $P_\Gamma$ as in Eq. (S4) around $P_\Gamma = \hat{P}_\Gamma$, i.e., $P_\Gamma = \hat{P}_\Gamma + \delta P_\Gamma$, will produce the equation below. For simplifying the notation we abbreviate, $P(\{n_q^{total}\})$ as $P_0$.

$$\delta S = 0 = \sum_\Gamma P_C(\ln P_C + \ln P_0)(\delta P_0) + \sum_\Gamma P_C(\delta P_0) \tag{S6}$$

The same variations in $P_\Gamma$ will yield the equation below from the constraint equation, Eq. (S5),

$$0 = \sum_\Gamma f_\Gamma P_C(\delta P_0) \tag{S7}$$

Therefore by solving for $\hat{P}(\{n_q^{total}\})$ using Eq. (S6) and Eq. (S7),

$$\hat{P}(\{n_q^{total}\}) = Z^{-1}Q_C \exp(-\lambda\sum_{\Gamma_C} f_\Gamma P_C) \tag{S8}$$

where, $\ln Q_C = -\sum_{\Gamma_C} P_C \ln P_C$ and $\Gamma_C$ describes the stochastic trajectory ($\{n_j\}$, $t_n$ ; $\{n_j\}$, $t_{n-1}$ ; $\{n_j\}$, $t_{n-2}$ ;….; $\{n_j\}$, $t_1$;$\{n_j\}$, $t_0$) with a fixed specific total protein abundances, $\{n_q^{total}\}$. Therefore, summation over $\Gamma_C$ essentially denotes averages over variations of stochastic trajectories due to intrinsic noise fluctuations. λ is the Lagrange multipliers, which is determined by substituting Eq. (S8) in constraint equation (Eq. (S5)). *Z* is the partition function. In deriving Eq. (S8), we also used the fact, $\sum_{\Gamma_C} P_C = 1$. Eq. (S8) can be easily generalized to include additional constraints describing average values and higher order moments of variables (such as τ, s, and p) describing chemotactic responses in single E.

coli cells. The conditional probability, $P_C$, and thus $Q_C$, in principle, can be calculated by solving the Master Eq. (S3) for the E. coli chemotaxis signaling models (MBL, BL or FT) we considered. We used an approximate numerical scheme to evaluate $Q_C$ in our simulations (see section B). However, in general, it is possible that the underlying biochemical reactions in a model are not known and consequently $P_C$ cannot be evaluated by solving the Master Equation in Eq. (S3). In similar situations, when it is possible to do repeated experiments, constraints on $\ln Q_C$ have been imposed to infer parameter distributions(21-23). Though we do not face this issue for the models in E. coli chemotaxis where the signaling reactions are very well characterized, the same formalism developed above can be used in signaling systems where the intrinsic noise fluctuations are not well characterized. In such cases, the distribution of total protein abundances can be evaluated by extending Eq. (S8) with additional Lagrange multipliers incorporating constraints for $\ln Q_C$. However, imposition of constraints on $\ln Q_C$ would require data available from repeated experiments probing signaling kinetics on the same individual cell or when independent signaling modules exist in the same cell which can be assayed simultaneously(24).

(B) *Calculation of $\hat{P}(\{n_q^{\text{total}}\})$ in simulations when the chemotactic responses are constrained*

We first create *a priori* distribution of the protein abundances by choosing total protein abundances at $t^{\text{initial}} = -800000$ s from a uniform distribution $U(0, U_H)$, where $U_H$ is chosen to be roughly 10 times larger than the experimentally measured mean abundance of the corresponding chemotactic protein (9). This step essentially creates a cell population where an individual cell is assigned with a set of total protein abundances chosen randomly from the above uniform distribution. We used up to 70,000 samples (or cells) drawn from the above uniform distribution. Then following the procedure described in section (II) we simulate the chemotactic signaling kinetics in individual cells. In the samples we used in our simulations, each *E. coli* cell produces a unique chemotactic response kinetics or a stochastic trajectory describing time evolution of abundances of signaling proteins, therefore we identify each trajectory by the single cell that generated it and vice versa. Thus the simulations generate a set of stochastic trajectories $\{\Gamma\}$ uniquely associated with a set of single cells. From this set of stochastic trajectories, $\{\Gamma\}$, we evaluated the distribution in Eq. (4). The summation over $\Gamma_C$ is carried out by using the unique association of a stochastic trajectory with a single cell in our simulations as this implies that for a particular stochastic trajectory, $\Gamma'_C = \{\{n'_j\}, t_n ; \{n'_j\}, t_{n-1} ; \ldots; \{n'_j\}, t_0\}$ associated with a fixed set of total protein concentrations $\{ n'^{\text{total}}_q\}$, $P_C = P(\{n'_j\}, t_n ; \{n'_j\}, t_{n-1} ; \ldots; \{n'_j\}, t_0 \mid \{n'^{\text{total}}_q\})$ is either equal to 1 (when $\Gamma'_C$ or the corresponding single cell is present in the samples we considered) or 0 (when $\Gamma'_C$ is absent in the samples). Thus, $\sum_{\Gamma_C} P_C \ln P_C = 0$ and

$$\sum_{\Gamma_C} f_\Gamma P_C = \sum_{\substack{\Gamma'_C \\ \text{or the trajectories present in the sample}}} f_{\Gamma'_C}\,.$$

It is in principle possible that when a very large number of cells are present, two different single cells could produce the same stochastic trajectory and for such cases, $\sum_{\Gamma_C} P_C \ln P_C$ will not vanish and $\sum_{\Gamma_C} f_\Gamma P_C$ will contain averages over multiple trajectories. However, occurrences of such events (e.g., the presence of pairs of identical stochastic trajectories) appear to be extremely rare for the rate constants and the ranges of the protein abundances we considered. We further tested this approximation by considering deterministic chemotactic signaling kinetics where the kinetics of signaling protein abundances only depend on the total protein abundances (as the kinetic rates are fixed for each cell), therefore, $P_C$=1 when the deterministic kinetic trajectory of abundances of signaling proteins is present and $P_C$=0, otherwise. When we used the same a priori uniform distribution for protein abundances as our stochastic simulations for the deterministic chemotactic kinetics, the qualitative results of the inferred $\hat{P}(\{n_q^{\text{total}}\})$ did not change compared to the stochastic simulations (Fig. S11). The small differences in the correlations shown in Fig. S11 between the stochastic and the deterministic simulations show the dominance of extrinsic noise fluctuations over intrinsic noise fluctuations in determining variations of protein abundances. These results also demonstrate that associating a unique stochastic trajectory to a single cell is a good approximation for the calculation of $\hat{P}(\{n_q^{\text{total}}\})$ in the stochastic simulations. We used the Levenberg-Marquardt algorithm in MATLAB to solve the nonlinear equations involving the Lagrange's multipliers.

*Values of the constraints used when the chemotactic responses were constrained*: We have considered average values, variances, and covariances of the variables characterizing the chemotactic response in E. coli. The variables are i) the adaptation time $\tau$, ii) the precision of adaptation $s$ and iii) the percentage variation of CheY-P about an optimal value $p$.

(i) Constraints for $\tau$: The distribution of $\tau$ ($P(\tau)$) is measured in Ref. (25). We use the distribution for 100 μM of aspartate stimulation to calculate the average value and the variance of $\tau$. $P(\tau)$ can be approximated fairly well with a normal distribution ($\chi^2$ =

0.0024179), therefore, the higher moments beyond the second moment do not capture relevant information for τ. Therefore, we did not use any higher order moments to further constrain τ in the MaxEnt calculations.

(ii) Constraints for s: The cell population averaged value of s for wildtype RP437 strain was obtained from Alon et al (5). The experimental measurements for the variance or any higher order moments of *s* are not available from the literature. Therefore, we used an ad hoc small value (0.0005) for the variance of s. In order to study the sensitivity of the inferred $\hat{P}(\{n_q^{\text{total}}\})$ upon any change in the distribution of *s*, we investigated the change in $\hat{P}(\{n_q^{\text{total}}\})$ when a different value was used for the variance of *s* (Fig. S13).

(iii) Constraints for *p*: Study of the switching frequency of single E. coli cells for the transformed PS2001strains for different steady state concentrations of CheY-P was carried out by Cluzel et al (6). They found that a more than 30 % variation in CheY-P concentration about an optimal value leads to a dramatic decrease in the switching frequency. Since Ref. (6) used engineered cells, we do not have a direct access to the distribution of *p* for the wild type E. coli cells. We used a cell population averaged value of *p* at 20 with a standard deviation 5. We also varied the standard deviation to study the sensitivity of $\hat{P}(\{n_q^{\text{total}}\})$ on the shape of the distribution of p (Fig. S13).

(C) *Evaluation of* $\hat{P}(\{n_q^{\text{total}}\})$ *when total protein abundances are constrained*: We considered two cases: (a) cell population averaged total abundances of the chemotactic proteins were constrained. (b) The pair wise correlations between total abundances of chemotactic proteins were constrained along with the average values. We maximized the Shannon's Entropy,

$$S = -\sum_{\{n_q^{\text{total}}\}} P\left(\left\{n_q^{\text{total}}\right\}\right)\ln\left(P\left(\left\{n_q^{\text{total}}\right\}\right)\right) \qquad \text{(S9)}$$

for both types of constraints.

Constraint on the mean: We constrained average values of the total protein abundances, i.e.,

$$\sum_{\{n_{\tilde{q}}^{\text{total}}\}} n_q^{\text{total}} P\left(\left\{n_{\tilde{q}}^{\text{total}}\right\}\right) = \overline{n_q}^{\text{expt}} \qquad \text{(S10)}$$

,where, q = 1…6, represent the proteins Tar, CheA, CheY, CheR, CheB and CheZ, respectively. The distribution that maximizes Eq. (S9) in the presence of the constraints in Eq. (S10) is given by

$$\hat{P}\left(\left\{n_q^{\text{total}}\right\}\right) \propto \prod_q \exp\left(-n_q^{total} / \overline{n_q}^{\text{expt}}\right) \tag{S11}$$

We drew total protein abundances from an exponential distribution of the form in Eq. (S11) and then used stochastic signaling kinetics to generate the distributions of the variables characterizing chemotactic responses.

<u>Constraint on the mean values, variances and covariances of total protein abundances:</u>
Here the mean values (Eq. S10) as well as the pair-wise covaraince ($C_{qq'}$) of the total protein abundances were constrained, i.e.,

$$\sum_{\left\{n_{\tilde{q}}^{\text{total}}\right\}} n_q^{\text{total}} n_{q'}^{\text{total}} P\left(\left\{n_{\tilde{q}}^{\text{total}}\right\}\right) - \overline{n_q}^{\text{expt}} \overline{n_{q'}}^{\text{expt}} = C_{qq'} \tag{S12}$$

The $\hat{P}(\{n_q^{\text{total}}\})$ that maximizes S in Eq. (S10) in the presence of constraints in Eqs. (S11-S12) is given by,

$$P\left(\left\{n_q^{\text{total}}\right\}\right) \propto \exp\left(\sum_{q,q'} n_q^{\text{total}} \cdot \mathbf{M} \cdot n_{q'}^{\text{total}}\right) + \sum_q \lambda_q n_q^{\text{total}} \tag{S13}$$

,where, $\mathbf{M}$ is a $6 \times 6$ symmetric matrix containing 21 independent Lagrange multipliers and $\{\lambda_q\}$ are 6 separate Lagrange multipliers arising from constraints for the mean values. For BL model, $\mathbf{M}$ is a $5 \times 5$ matrix with 15 independent elements and there are 5 different $\lambda_q$ s.

We have represented the distribution in Eq. (S13) as a multivariate Gaussian distribution of the form

$$P\left(\left\{n_q^{\text{total}}\right\}\right) \propto \exp\left(-\frac{1}{2}\sum_{q,q'}\left(n_q^{total} - \overline{n_q}^{\text{expt}}\right) \cdot \mathbf{C}^{-1} \cdot \left(n_{q'}^{total} - \overline{n_{q'}}^{\text{expt}}\right)\right) \tag{S14}$$

where, $\mathbf{C}$ is the $6 \times 6$ covariance matrix in Eq (S12). Thus, the Lagrange multipliers in $\mathbf{M}$ and $\{\lambda_q\}$ are determined from the mean values and the $\mathbf{C}$ matrix obtained from the experiments or constraints.

Kollmann et al (1) measured covariances between CheZ, CheY and CheA, CheY pairs. Ref. (1) proposed a log-normal distribution for the describing the joint distribution for all the total protein abundances. The proposed log-normal distribution agrees well with the measured distributions in Kollmann et al., and, we found that when we used that distribution it reproduced the chemotactic responses reasonably well (Fig. S7 red plots). So, we used the values computed from the log-nornal distribution to constrain the covariances in our calculation.

According to Ref. (1)

$$n_q^{\text{total}} \approx \overline{n_q}^{\text{expt}} \left(\lambda r + 0.2\sqrt{\lambda} \cdot \xi_q\right), \tag{S15}$$

where $\lambda$ is the protein overexpression factor (should not be confused with $\lambda_q$), for the wild type RP437 strain $\lambda$=1. $\xi_q$ is a Gaussian variable with zero mean and unit variance and $r$ is a Log normal variable with unit mean. Using Eq. (S15) we calculated the **C** matrix as

$$C_{qq'} \approx 0.25 \times \overline{n}_q^{\text{expt}} \overline{n}_{q'}^{\text{expt}} + \rho \delta'_{qq} \tag{S16}$$

, where, ρ=0.04. We used the C matrix in Eq. (S16) to calculate the distribution in Eq. (S14). We simulated the distribution in Eq. (S16) by drawing random numbers from the multivariate Gaussian distribution given by Eq. (S14).

## IV. Evaluation of the minimal number of constraints required to reproduce the observed chemotactic response in E. coli

We drew intuition for constructing the case that will require the minimum number of constraints by considering the form of the C matrix in Eq. (S16). The C matrix in Eq. (S16) produces a strong Pearson correlation co-efficient of about 0.85 for any two pairs. When we approximate Eq. (S16) by,

$$C_{qq'} \approx 0.25 \times \overline{n}_q^{\text{expt}} \overline{n}_{q'}^{\text{expt}} \tag{S17}$$

then Pearson correlation co-efficient between protein pairs is exactly 1. The C matrix in Eq. (S17) is singular with one non-zero eigenvalue equal to $e_1 = 0.25 \times \sum_{q=1}^{6} (\overline{n}_q^{\text{expt}})^2$ , and the rest of the eigenvalues are zero, i.e., $e_2 = \cdots = e_6 = 0$. Using the eigenvectors of the **C** matrix in Eq. (S17) it can be shown that when any total protein abundance, (say $n_1^{\text{total}}$), follows a Gaussian distribution with a mean $\overline{n}_1^{\text{expt}}$ and a variance, $C_{11} = 0.25 \times (\overline{n}_1^{\text{expt}})^2$, and rest of the total protein abundances are related as follows:

$$\bar{n}_{q-1}^{\text{expt}}(n_q^{\text{total}} - \bar{n}_q^{\text{expt}}) = \bar{n}_q^{\text{expt}}(n_{q-1}^{\text{total}} - \bar{n}_{q-1}^{\text{expt}}) \qquad \text{(S18)}$$

, where, q=2..6, then the resulting distribution produces the C matrix in Eq. (S17). The relation in Eq. (S18) indicates a strong dependence between the proteins, e.g., protein 2 is regulated by protein 1, then protein 3 is regulated by protein 2, and so on. Thus, it indicated as long as, the proteins are connected to each other following a linear chain describing Eq. (S18), one could reproduce the correlation in Eq. (S17). The ordering of the proteins is not important as long as they are connected to the nearest neighbors. Next we checked how well the distribution of total protein abundances constructed using Eq. (S17) and mean values from Eq. (S10) can reproduce the chemotactic responses. We found the distribution reproduces the observed response reasonably well (Figs. S9-S10 maroon plots). Therefore, it indicates that the minimum number of constraints in this case would be composed of the mean values of the total protein abundances (6 in total) and the variance for one protein (say $C_{11}$).

Now we consider the C matrix in Eq. (S16), which is not a singular matrix, therefore, the strict equality between the protein abundances as given by Eq. (S18) is relaxed. However, drawing from the intuition developed from the above calculations, we constrain a set of pair correlations that will connect the proteins in a linear chain, e.g., the covariances between, Tar-CheA, CheA-CheY, CheY-CheR, CheR-CheB, and CheB-CheZ. We also constrained the mean values of the proteins as well. We then evaluated the corresponding multivariate Gaussian distribution for $\hat{P}(\{n_q^{\text{total}}\})$ that generates the constrained covariances and average values, and then use the distribution to generate the chemotactic responses (Fig. S7 and Fig S9 blue plots, Fig. S10 both blue and brown plots). We have studied two values of $\rho$ in Eq. (S16), 0.01 and 0.04, respectively. For $\rho = 0.01$, the nearest neighbor correlation coefficients are 0.95 while for $\rho = 0.04$, the nearest neighbor correlation coefficients are 0.83. We found that for $\rho = 0.01$, the distributions of $\tau$, s, and, p are in reasonable agreement with the experiments (Figs. S9-S10). For $\rho = 0.04$, the distribution for *p* becomes slightly broader. Thus, as long as we impose a strong correlation between the protein abundances by linking protein abundances in a linear chain, the resulting $\hat{P}(\{n_q^{\text{total}}\})$ reproduces the chemotactic responses reasonably well. Next we further tested the sensitivity of the linked network by removing a link, e.g., we constrained the covariances between, Tar-CheA, CheA-CheY, CheY-CheR, and CheR-CheB were constrained but the covariance between CheB-CheZ *was not constrained*.

Results for the one deleted link case: In this case, instead of 17 (6 mean values, 6 variances, and 5 covariances) constraints as in the linked case we will have 16 constraints (6 mean values, 6 variances, and 4 covariances). Both for $\rho = 0.01$ or $\rho = 0.04$, deletion of a link produces a large change in the distribution of *p*, making it broader (Figs. S7 and S9 indigo plots) suggesting a larger fraction of the E. coli cell will possess non-functional flagellar motors. Thus, this test further supports the mechanism, that as long as the total protein abundances are strongly linked by pair-correlations through the constraints, the resulting population will capture the observed chemotactic response reasonably well.

## V. Calculation of the $\chi^2$ distance to quantify comparison between the chemotactic responses obtained from inferred MaxEnt distributions and the experiments

The experimental values are shown in red in Table S5A. The values that we estimated are shown in green in that table. The standard errors ($\sigma$) for the assumed mean values are assumed to be of the same order as the mean values. We calculated the chi-square ($\chi^2$) given by,

$$\chi^2 = \sum_{i=1}^{5} \left( \frac{\overline{O^i_{\text{inferred}}} - \overline{O^i_{\text{expt}}}}{\sigma_{\overline{O^i_{\text{expt}}}}} \right)^2 \qquad \text{(S19)}$$

where the overbar indicates average values, and σ indicates the variance. The $O^i$ is the $i^{\text{th}}$ variable describing the chemotactic responses and $\sigma_{\overline{O^i_{\text{expt}}}}$ is the standard error in the variable. The subscripts "inferred" and "expt" denote the values obtained using the inferred distribution and the experimental measurements, respectively. Smaller values of $\chi^2$ indicate a better agreement. The $\chi^2$ values of the inferred distributions are shown in Table S5B-S5D.

**Table S1A: Average values of protein abundances used in the model**

| Species | Concentration (μM) |
|---|---|
| Receptor Total (T=Tar + Tsr) | 17.8 |
| CheA | 5.3 |
| CheB | 0.28 |
| CheR | 0.16 |
| CheY | 9.7 |
| CheZ (MBL and FT) | 3.8 |

Cell volume is 1.4 fl. The average concentrations of the chemotactic proteins for the RP437 strain have been taken from (9).

For a reaction line $\text{Tar}_m^{\text{inact}} \underset{k_{-1}}{\overset{k_1}{\rightleftharpoons}} \text{Tar}_m^A$, we have defined $a_m = k_1/(k_1 + k_{-1})$ such that in the steady state $\text{Tar}_m^A = a_m \cdot \text{Tar}_m^{\text{Total}}$. In the absence of the ligands we have assumed that $a_m = m / m_{\mathrm{Max}}$. In the presence of 100 μM aspartate, we assume that all the Tar molecules are ligand bound. The values of $a_m$ used are shown below. We chose these values in order to match the average adaptation time in our model to the one quoted in the experiment (25).

**Table S1B: The activation probabilities**

| **Reactions** | **a(L = 0 μM)** | **a(L =100 μM)** |
|---|---|---|
| $T_1^{inact} \rightarrow T_1^A$ | 0.125 | 0.0014 |
| $T_2^{inact} \rightarrow T_2^A$ | 0.5 | 0.0115 |
| $T_3^{inact} \rightarrow T_3^A$ | 0.874 | 0.056 |
| $T_4^{inact} \rightarrow T_4^A$ | 1.0 | 1.0 |

**Table S2A: Reactions for MBL model**

| **Reactions** | **Michaelis-Menten Constant** | **Rates** |
|---|---|---|
| $T_m + \text{CheR} \underset{k_{-1}}{\overset{k_1}{\rightleftarrows}} T_m\text{-CheR} \xrightarrow{k_r} T_{m+1} + \text{CheR}$ | $K_r=(k_{-1} + k_r)/k_1 = 0.099$ μM | $k_r = 0.39\ s^{-1}$ |
| $T_m^A + \text{CheB-P} \underset{k_{-1}}{\overset{k_1}{\rightleftarrows}} T_m^A\text{-CheB-P} \xrightarrow{k_b} T_{m-1}^A + \text{CheB-P}$ | $K_b=(k_{-1} + k_b)/k_1 = 2.5$ μM | $k_b = 6.3\ s^{-1}$ |
| $\text{CheA} + T^A \rightarrow \text{CheA-P} + T^A$ | | $k_p = 50\ \mu M^{-1}s^{-1}$ |
| $\text{CheB} + \text{CheA-P} \rightarrow \text{CheB-P} + \text{CheA}$ | | $a_b = 3\ \mu M^{-1}s^{-1}$ |
| $\text{CheB-P} \rightarrow \text{CheB}$ | | $d_b = 1.0\ s^{-1}$ |
| $\text{CheY} + \text{CheA-P} \rightarrow \text{CheY-P} + \text{CheA}$ | | $a_y = 100\ \mu M^{-1}s^{-1}$ |
| $\text{CheY-P} \rightarrow \text{CheY}$ | | $d_y = 0.1\ s^{-1}$ |
| $\text{CheY-P} + \text{CheZ} \rightarrow \text{CheY} + \text{CheZ}$ | | $k_Z = 7.8\ \mu M^{-1}s^{-1}$ |

The rate constants are taken from Kollmann et al.

**Table S2B: Reactions for FT model**

| **Reactions** | **Michaelis-Menten Constant** | **Rates** |
|---|---|---|
| $T_m + \text{CheR} \underset{k_{-1}}{\overset{k_1}{\rightleftarrows}} T_m\text{-CheR} \xrightarrow{k_r} T_{m+1} + \text{CheR}$ | $K_r=(k_{-1} + k_r)/k_1 = 0.099$ μM | $k_r = 0.39\ s^{-1}$ |
| $T_m + \text{CheB-P} \underset{k_{-1}}{\overset{k_1}{\rightleftarrows}} T_m\text{-CheB-P} \xrightarrow{k_b} T_{m-1} + \text{CheB-P}$ | $K_b=(k_{-1} + k_b)/k_1 = 2.5$ μM | $k_b = 6.3\ s^{-1}$ |
| $\text{CheA} + T^A \rightarrow \text{CheA-P} + T^A$ | | $k_p = 50\ \mu M^{-1}s^{-1}$ |
| $\text{CheB} + \text{CheA-P} \rightarrow \text{CheB-P} + \text{CheA}$ | | $a_b = 3\ \mu M^{-1}s^{-1}$ |
| CheB-P $\rightarrow$ CheB | | $d_b = 1.0\ s^{-1}$ |
| CheY + CheA-P $\rightarrow$ CheY-P + CheA | | $a_y = 100\ \mu M^{-1}s^{-1}$ |
| CheY-P $\rightarrow$ CheY | | $d_y = 0.1\ s^{-1}$ |
| CheY-P + CheZ $\rightarrow$ CheY + CheZ | | $k_Z = 7.8\ \mu M^{-1}s^{-1}$ |

**Table S2C: Reactions for BL model**

| **Reactions** | **Michaelis-Menten Constant** | **Rates** |
|---|---|---|
| $T_m + \text{CheR} \underset{k_{-1}}{\overset{k_1}{\rightleftharpoons}} T_m\text{-CheR} \xrightarrow{k_r} T_{m+1} + \text{CheR}$ | $K_r=(k_{-1} + k_r)/k_1 = 0.099\ \mu M$ | $k_r = 0.39\ s^{-1}$ |
| $T_m^A + \text{CheB-P} \underset{k_{-1}}{\overset{k_1}{\rightleftharpoons}} T_m^A\text{-CheB-P} \xrightarrow{k_b} T_{m-1}^A + \text{CheB-P}$ | $K_b=(k_{-1} + k_b)/k_1 = 2.5\ \mu M$ | $k_b = 6.3\ s^{-1}$ † |
| CheA + $T^A$ $\rightarrow$ CheA-P + $T^A$ | | $k_p = 50\ \mu M^{-1}s^{-1}$ |
| CheB + CheA-P $\rightarrow$ CheB-P + CheA | | $a_b = 3\ \mu M^{-1}s^{-1}$ |
| CheB-P $\rightarrow$ CheB | | $d_b = 1.0\ s^{-1}$ |
| CheY + CheA-P $\rightarrow$ CheY-P + CheA | | $a_y = 100\ \mu M^{-1}s^{-1}$ |
| CheY-P $\rightarrow$ CheY | | $d_y = 30\ s^{-1}$ |

† The $K_b$ and $k_b$ values used by Kollmann et al for the BL model are 16 μM and 16 $s^{-1}$ respectively. We have used the same rate constants for all the three models to facilitate comparison across them.

**Table S3: Comparison of the average values of protein abundances generated by the MaxEnt distribution when different combinations of the chemotactic outputs ($\tau$, $\tau^2$, $s$, $s^2$, $p$, $p^2$) are constrained to the cell population averages measured in experiments**.

**A: Experiments (9)**

| | **Tar+Tsr molecules/ cell** | **CheA molecules/ cell** | **CheY molecules/ cell** | **CheR molecules/ cell** | **CheB molecules/ cell** | **CheZ molecules/ cell** |
|---|---|---|---|---|---|---|
| | 15000 ± 1700 | 4452 ± 920 | 8148 ± 310 | 140 ± 10 | 240 ± 10 | 3200 ± 90 |

**B: The values quoted are for MBL model.**

| Constraints | Tar+Tsr molecules / cell | CheA molecules / cell | CheY molecules / cell | CheR molecules / cell | CheB molecules / cell | CheZ molecules / cell |
|---|---|---|---|---|---|---|
| $\overline{\tau} = 245$ s | 46698 | 14584 | 24404 | 786 | 466 | 10198 |
| $\overline{p} = 20$ | 46318 | 13025 | 21015 | 667 | 547 | 12543 |
| $\overline{s} = 0.02$ | 48933 | 15409 | 23570 | 838 | 465 | 9566 |
| $\overline{\tau} = 245$ s<br>$\overline{s} = 0.02$<br>$\overline{p} = 20$<br>$\overline{\tau^2} = 62323$ s$^2$<br>$\overline{p^2} = 425$ | 59991 | 12693 | 24670 | 504 | 538 | 12909 |
| $\overline{s} = 0.02$<br>$\overline{p} = 20$<br>$\overline{\tau^2} = 62323$ s$^2$<br>$\overline{p^2} = 425$<br>$\overline{s^2} = 0.0005$ | 60033 | 13023 | 24630 | 491 | 545 | 13051 |

**Table S4: Pearson correlations between protein abundances.**
Average values of $s$, $p$, $\tau$, $\tau^2$, and $p^2$ were constrained in the MaxEnt calculation.

| Protein pair | Sign of Pearson correlation from the MaxEnt calculation | | Single Cell Experiment (Fig. 2, Kollmann et al., Nature, 2005) | In vitro assay of cloned gene pairs (Fig. 1, Lovdok et al. , PloS Biol., 2009) (26) | Bioinformatic Analysis of pairwise gene co-occurrence (Table 1, Lovdok et al. , PloS Biol., 2009) (26) |
|---|---|---|---|---|---|
| | MBL | FT | | | |
| Tar-CheA (^) | - | - | | | + (10.5) |
| Tar-CheY | - | + | | | + (1.9*) |
| Tar-CheR | + | - | | | + (16.8) |
| Tar-CheB | + | + | | | +(2.3*) |
| Tar-CheZ | + | + | | | +(<1*) |
| CheA-CheY (^) | - | - | + | | +(15.7) |
| CheA-CheR (^) | - | + | | | +(3.9*) |
| CheA-CheB (^) | + | - | | | +(15.1) |
| CheA-CheZ (^) | + | - | | | +(32.5) |
| CheY-CheR | + | - | | | +(1.9*) |
| CheY-CheB | + | + | | + | +(15.0) |
| CheY-CheZ | + | + | + | + | +(90.0) |
| CheR-CheB | - | - | | + | +(28.6) |
| CheR-CheZ | + | - | | | 0 |
| CheB-CheZ | - | + | | | +(<1*) |

* Indicates small pair-wise occurrence in bacterial genomes.
^ Indicates protein pairs encoded by genes in the same operon (*meche*).
Signs marked in blue and red indicate agreement and disagreement, respectively, between MaxEnt predictions and experiments.
Rows shaded with blue denote positive correlations between protein abundances predicted by the MaxEnt approach for the MBL model.

**Table S5: Table for the $\chi^2$ for the average chemotactic outputs obtained from the distributions of the τ, p, and s in Fig S7-S8 and the experimentally observed values.**

**A: Values obtained from the Experimental measurements**

| | $\overline{\tau}$ sec | $\overline{\tau^2}$ $sec^2$ | $\sigma_{\overline{\tau}}$ sec | $\sigma_{\overline{\tau^2}}$ $sec^2$ | $\overline{s}$ | $\overline{p}$ | $\overline{p^2}$ | $\sigma_{\overline{s}}$ | $\sigma_{\overline{p}}$ | $\sigma_{\overline{p^2}}$ |
|---|---|---|---|---|---|---|---|---|---|---|
| Experiments | 250.4 | 65092 | 18.6 | 9205 | 0.02 | 20 | 425 | 0.01 | 10 | 250 |

**B: Multivariate Gaussian (constrained means, variances and correlations, Fig 4 main text and Fig S8)**

| | $\overline{\tau}$ sec | $\overline{\tau^2}$ $sec^2$ | $\overline{s}$ | $\overline{p}$ | $\overline{p^2}$ | $\chi^2$ |
|---|---|---|---|---|---|---|
| MBL (Total 27 constraints) | 277 | 94338 | 0.03 | 24 | 1026 | 20.0 |
| BL (Total 21 constraints) | 226 | 59272 | 0.05 | 49 | 4045 | 229 |
| FT (Total | 226 | 65487 | 0.3 | 31 | 1500 | 829 |
| 27 constraints) | | | | | | |

**C: $\chi^2$ for the average chemotactic outputs obtained from the distributions of the phenotypes in Fig S7 from the experiments.**

| MaxEnt distributions for MBL model | $\bar{\tau}$ sec | $\overline{\tau^2}$ $sec^2$ | $\bar{s}$ | $\bar{p}$ | $\overline{p^2}$ | $\chi^2$ |
|---|---|---|---|---|---|---|
| Exponential (6 constraints) | 401 | 366248 | 0.09 | 111 | 35228 | 20641 |
| Gaussian (27 constraints) | 277 | 94338 | 0.03 | 24 | 1026 | 20.0 |
| Nearest Neigbor (17 constraints, ρ = 0.04) | 280 | 105020 | 0.04 | 44 | 3821 | 215 |
| Nearest Neigbor (17 constraints, ρ = 0.01) | 270 | 82978 | 0.03 | 24 | 1048 | 13 |
| One Deleted link (16 constraints, ρ = 0.04) | 284 | 113332 | 0.04 | 65 | 9477 | 1366 |
| One Deleted link (16 constraints, ρ = 0.01) | 278 | 99564 | 0.04 | 61 | 8125 | 984 |

**D: $\chi^2$ for the average chemotactic outputs obtained from the distributions of the phenotypes when the protein numbers are drawn from log-normal distribution** (Kollmann et al (1)).

| MaxEnt distributions for MBL model | $\overline{\tau}$ sec | $\overline{\tau^2}$ sec$^2$ | $\overline{s}$ | $\overline{p}$ | $\overline{p^2}$ | $\chi^2$ |
|---|---|---|---|---|---|---|
| Log-normal distribution | 264 | 80620 | 0.06 | 26 | 1158 | 30 |

**Table S6**: **The $\chi^2$ of the first and second order cumulants of CheY, CheZ and CheA obtained from the MaxEnt distribution of the MBL model when $\tau$, $\tau^2$, s, $p$, $p^2$ are constrained to the experimental value.**

| | **Average$_{\mathbf{MBL}}$** $\overline{O^i_{dist}}$ | **Average$_{\mathbf{expt}}$** $\overline{O^i_{\text{expt}}}$ | **Error$_{\mathbf{expt}}$** $\sigma_{\overline{O^i_{\text{expt}}}}$ | $\chi^2 = \sum_{i=1}^{8}\left(\frac{\overline{O^i_{\text{inferred}}} - \overline{O^i_{\text{expt}}}}{\sigma_{\overline{O^i_{\text{expt}}}}}\right)^2$ |
|---|---|---|---|---|
| $\overline{\text{CheY}}$ | 24670 | 8148 | 507 | |
| $\overline{\text{CheZ}}$ | 12909 | 3192 | 223 | |
| $\overline{\text{CheA}}$ | 12693 | 4452 | 233 | 32500 |
| $\overline{\text{CheY}^2}$ | 7.80057e+08 | 7.79465e+07 | 9.1372e+06 | |
| $\overline{\text{CheZ}^2}$ | 1.89109e+08 | 1.2914e+07 | 2.0356e+06 | |
| $\overline{\text{CheA}^2}$ | 2.25765e+08 | 2.22611e+07 | 2.42987e+06 | |
| $\overline{\text{CheY}\bullet\text{CheZ}}$ | 3.40111e+08 | 3.19421e+07 | 5.3621e+06 | |
| $\overline{\text{CheY}\bullet\text{CheA}}$ | 3.00493e+08 | 3.92766e+07 | 3.86588e+06 | |

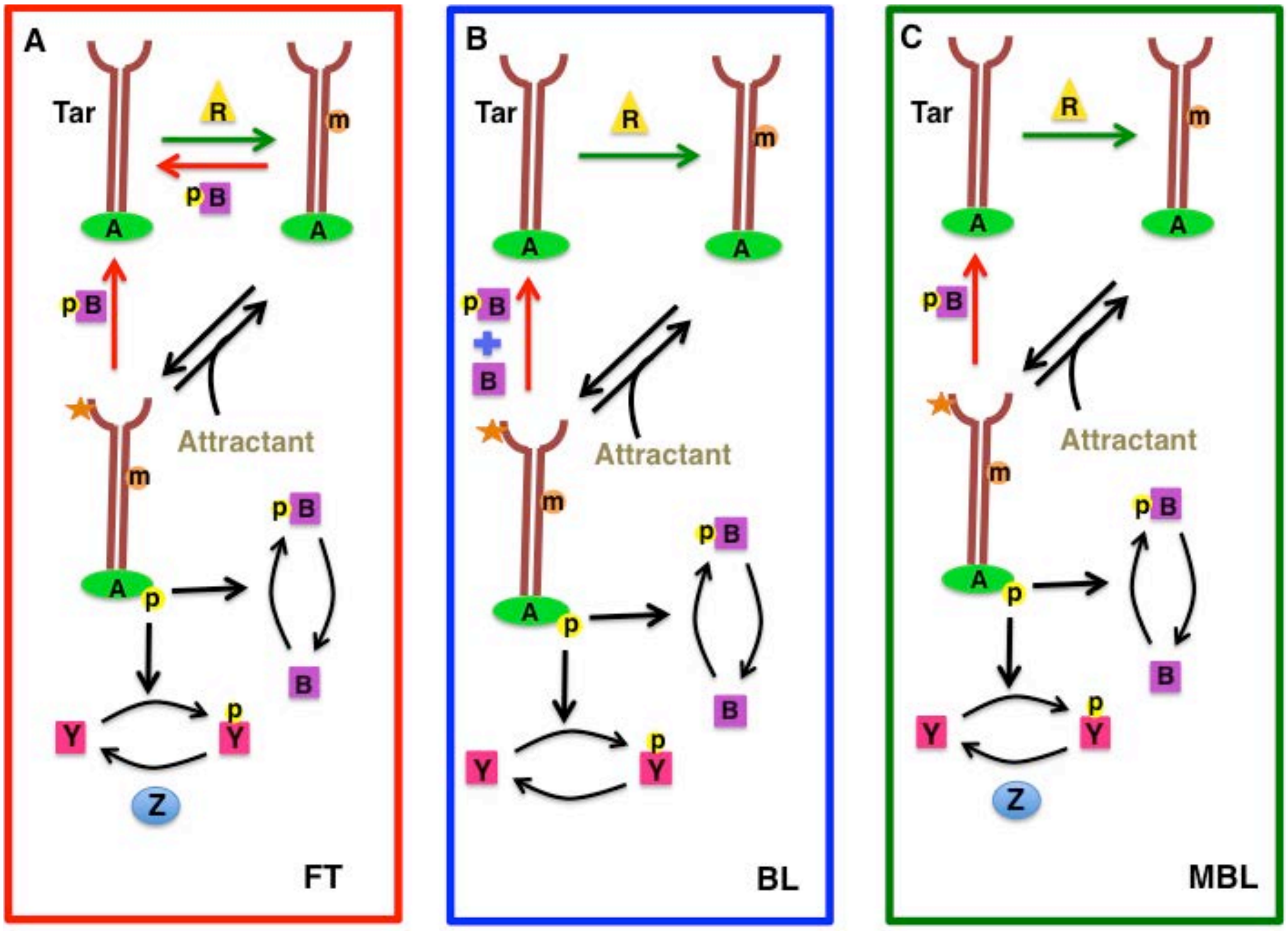

**Fig. S1 Three models for *E. coli* chemotaxis. (A)** The Fine tuned model (FT), originally proposed by Knox et al and Hauri and Ross, shows robust adaptation only for a small region in the parameter space. In this model the aspartate receptors (Tar) can shuttle between an active (denoted by the orange star) and an inactive conformation. The probability of the receptors to be in an active conformation depends on their state of methylation. Pair of enzymes namely CheR (R) and CheB-P (B) add and remove methyl groups from the receptors. CheB-P can demethylate receptors regardless of *their state of activity*. The Tar receptors form complexes with a kinase CheA (A). CheA can phosphorylate itself with a rate proportional to the abundance of the active Tar. Phosphorylated CheA can transfer its phosphoryl group ($PO_4$) either to the kinase CheB, rendering it capable of demethylation, or to another response regulatory protein called CheY (Y). Upon receiving the phosphoryl group from CheA, CheY renders itself active. Active form of CheY (CheY-P) diffuses across the cell and binds to the flagella motors causing them to tumble. The phosphatase CheZ (Z) de-activates the active form CheY. **(B)** Barkai Leibler model (BL) was put forward to explain robust adaptation in the chemotactic network of bacterial *E. coli*. Unlike FT model, CheB, both the unphosphorylated and the phosphorylated form, demethylates only the *active* Tar receptors. This model also lacks the phosphatase CheZ. **(C)** MBL model proposed by Kollmann et al., albeit for two differences, is exactly similar to the BL model. The differences are i) Unlike $CheB^T$ (CheB-P and CheB), only the *phosphorylated* form of CheB can de-methylate the active receptors in this model. ii) The dephosphorylation of the active CheY is done by the phosphatase called CheZ (Z). The presence of this phosphatase makes the steady state of CheY-P relatively robust to the concerted over expression of the chemotactic proteins. We use the figure (Fig. 1) from Mukherjee et al, (8) here for the convenience of the readers.

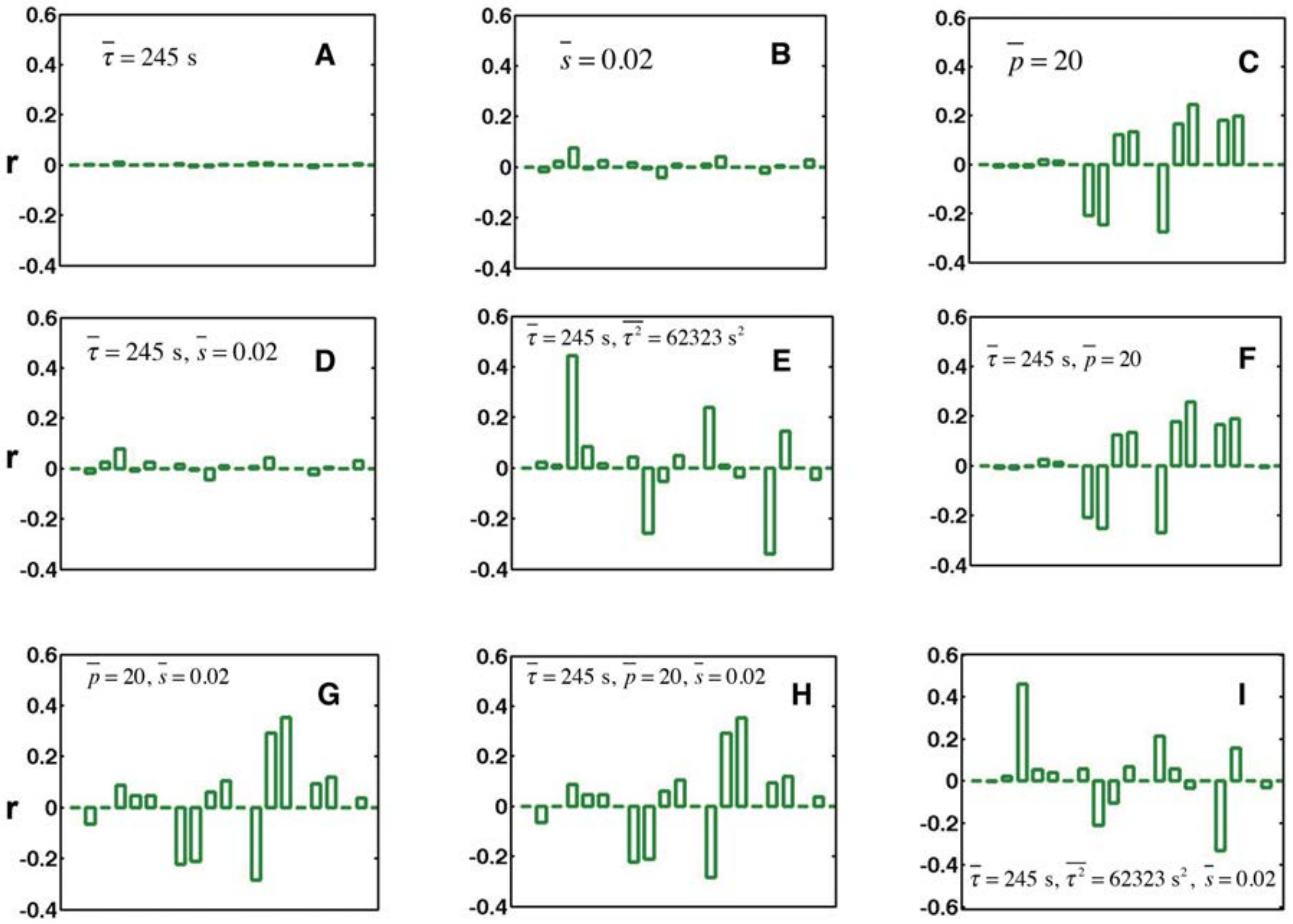

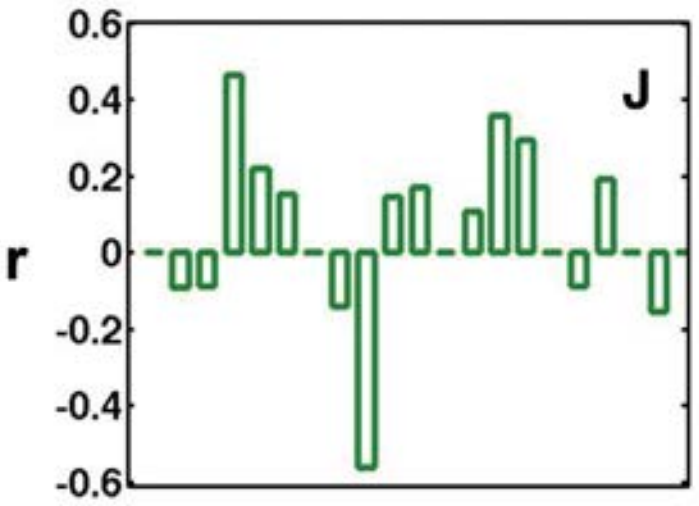

$\bar{\tau} = 245\ \text{s},\ \overline{\tau^2} = 62323\ \text{s}^2,\ \bar{p} = 20$

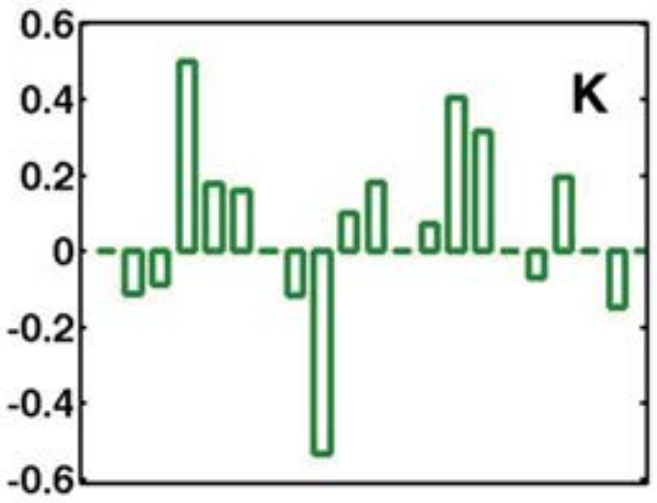

$\bar{\tau} = 245\ \text{s},\ \overline{\tau^2} = 62323\ \text{s}^2,\ \bar{p} = 20,\ \bar{s} = 0.02$

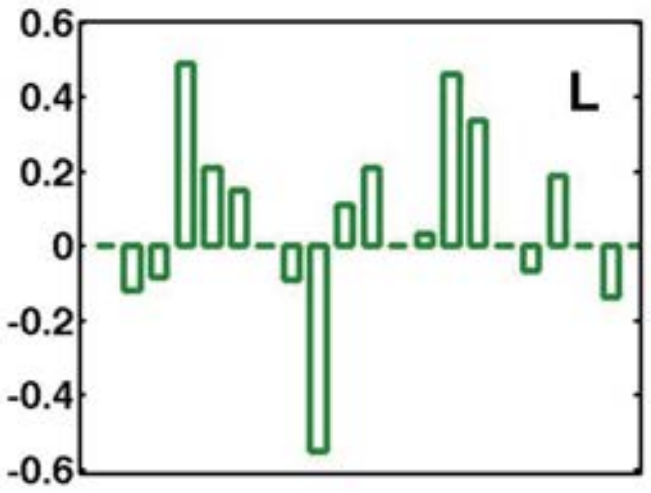

$\bar{\tau} = 245\ \text{s},\ \overline{\tau^2} = 62323\ \text{s}^2,\ \bar{p} = 20$
$\overline{p^2} = 425,\ \bar{s} = 0.02$

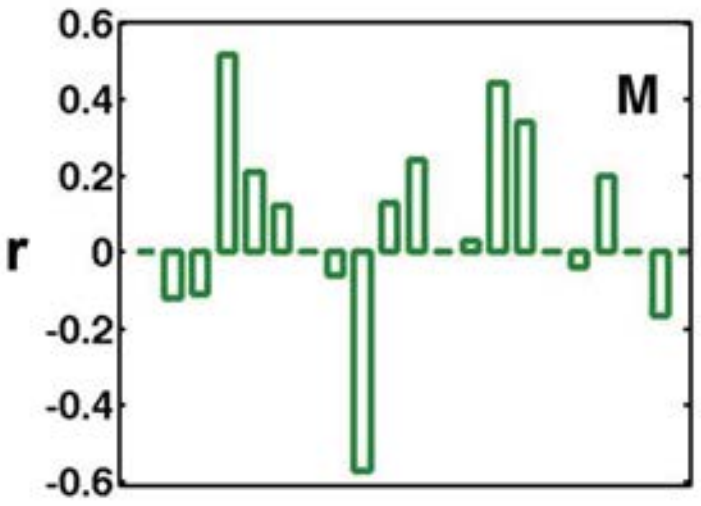

$\bar{\tau} = 245\ \text{s},\ \overline{\tau^2} = 62323\ \text{s}^2,\ \bar{p} = 20$
$\overline{p^2} = 425,\ \bar{s} = 0.02,\ \overline{s^2} = 0.0005$

**Fig S2: Pearson correlations are shown for the MBL for different combinations of the chemotactic constraints.** The green bars from left to right represent the correlation co-effiecients between Tar-Tar, Tar-CheA, Tar-CheY, Tar-CheR, Tar-CheB, Tar-CheZ, CheA-CheA, CheA-CheY, CheA-CheR, CheA-CheB, CheA-CheZ, CheY-CheY, CheY-CheR, CheY-CheB, CheY-CheZ, CheR-CheR, CheR-CheB, CheR-CheZ, CheB-CheB, CheB-CheZ and CheZ-CheZ respectively **(A)** The average value of the adaptation time is constrained to the experimental average of 245 s. **(B)** The average value of the precision of adaptation is constrained to the experimental value of 0.02. **(C)** The average value of the percentage variation in the steady states of CheY-P is constrained to 20%. **(D)** The average value of the adaptation time and the precision of adaptation are both constrained to the experimental values of 245 s and 0.02. **(E)** The average and the second moment of the adaptation time are constrained to 245s and 62323s$^2$ respectively. **(F)** The average value of the adaptation time and the percentage variation in the steady state of CheY-P are constrained to 245s and 20 %. **(G)** The steady state variation of the CheY-P protein and the precision of adaptation are constrained to 20% and 0.02 respectively. **(H)** All the three chemotactic outputs, the adaptation time, precision of adaptation and the variation in the steady state of CheY-P from the optimal value are constrained to 245s, 0.02 and 20% respectively. **(I)** The average adaptation time, the second moment of the adaptation time and the precision of adaptation are constrained to 245 s, 62323 s$^2$ and 0.02 respectively. **(J)** The average value of the adaptation time, the second moment of the adaptation time and the percentage variation of the steady state of CheY-P are constrained to 245 s, 62323 s$^2$ and 20% respectively. **(K)** The average value of the adaptation time, its second moment, the precision of adaptation and the variation in the steady state of CheY-P are constrained to 245 s, 62323 $s^2$ , 0.02 and 20%. **(L)** The adaptation time, the second moment of the adaptation time, the precision of adaptation, the variation in the steady state of CheY-P and its second moment are all constrained to 24s s, 62323 $s^2$, 0.02, 20% and 425 respectively. **(M)** On top of the 5 constraints in S5 (L) we added an extra constraint of 0.0005 on the second moment of the precision of adaptation. We show the difference, $r = r_{MaxEnt} - r_{uni}$ for different protein pairs, where, $r_{\text{uni}}$ and $r_{\text{MaxEnt}}$ refer to the Pearson correlation calculated for the *a priori* uniform and the MaxEnt distribution, respectively. The number of samples for all the cases is 70,000.

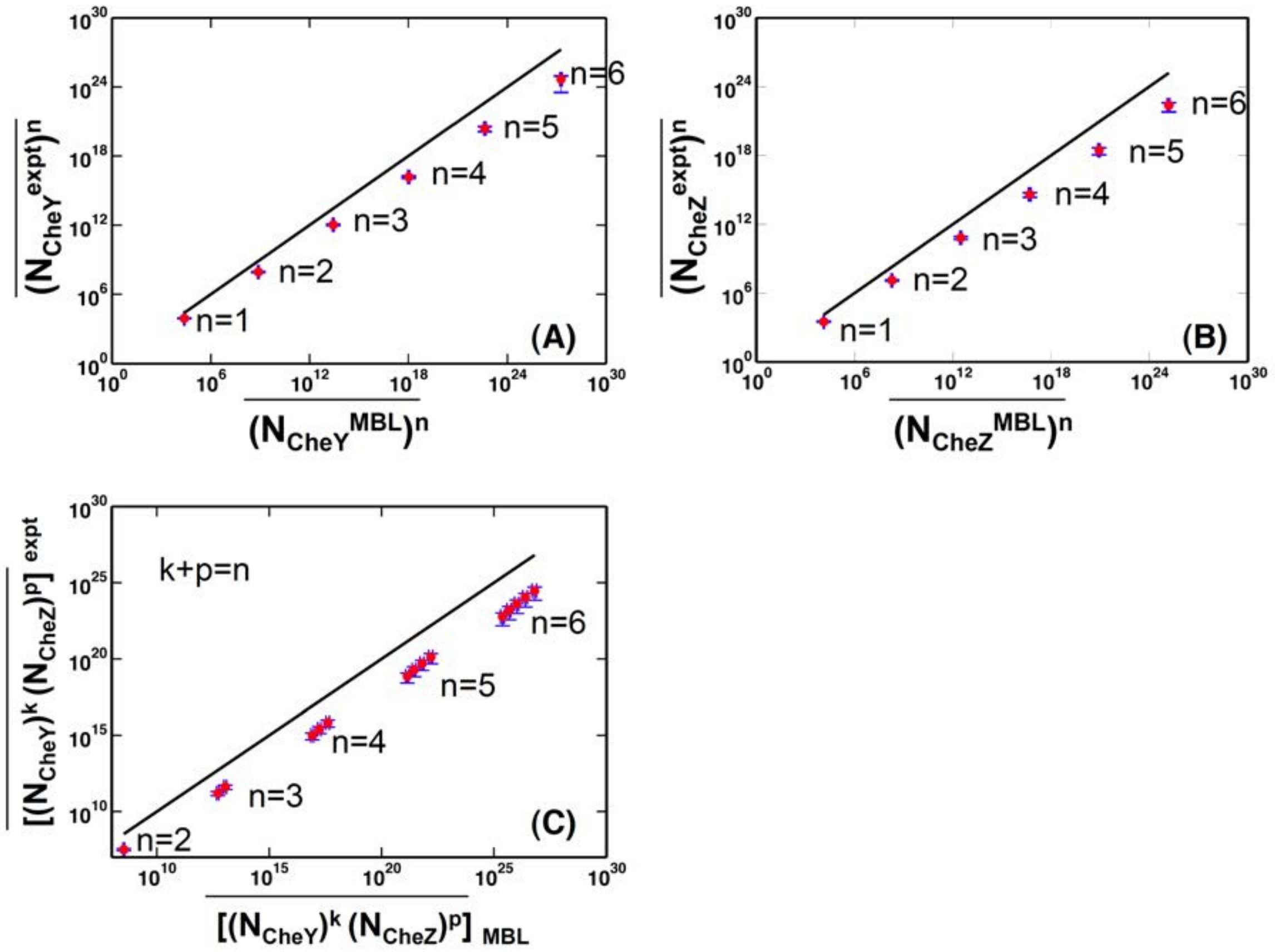

**Fig S3: A comparison of individual and cross moments of CheY and CheZ when six chemotactic outputs are constrained in the MBL model.** We calculate all the moments up to the sixth order. The constraints imposed for all the plots above are

$\overline{\tau} = 245$ s, $\overline{\tau^2} = 62323$ $s^2$, $\overline{s} = 0.02$, $\overline{p} = 20$, $\overline{p^2} = 425$, $\overline{s^2} = 0.0005$ .

**(A)** Comparison of the moments of CheY abundances calculated from the MaxEnt distribution with the data from single cell experiments in Ref.(1). The y=x line (solid black) is shown for comparison. **(B)** Similar comparison as in (A) for CheZ abundances. **(C)** Similar comparison as in (A) for the joint distribution of CheZ and CheY abundances. The y=x line (solid black) is shown for comparison.

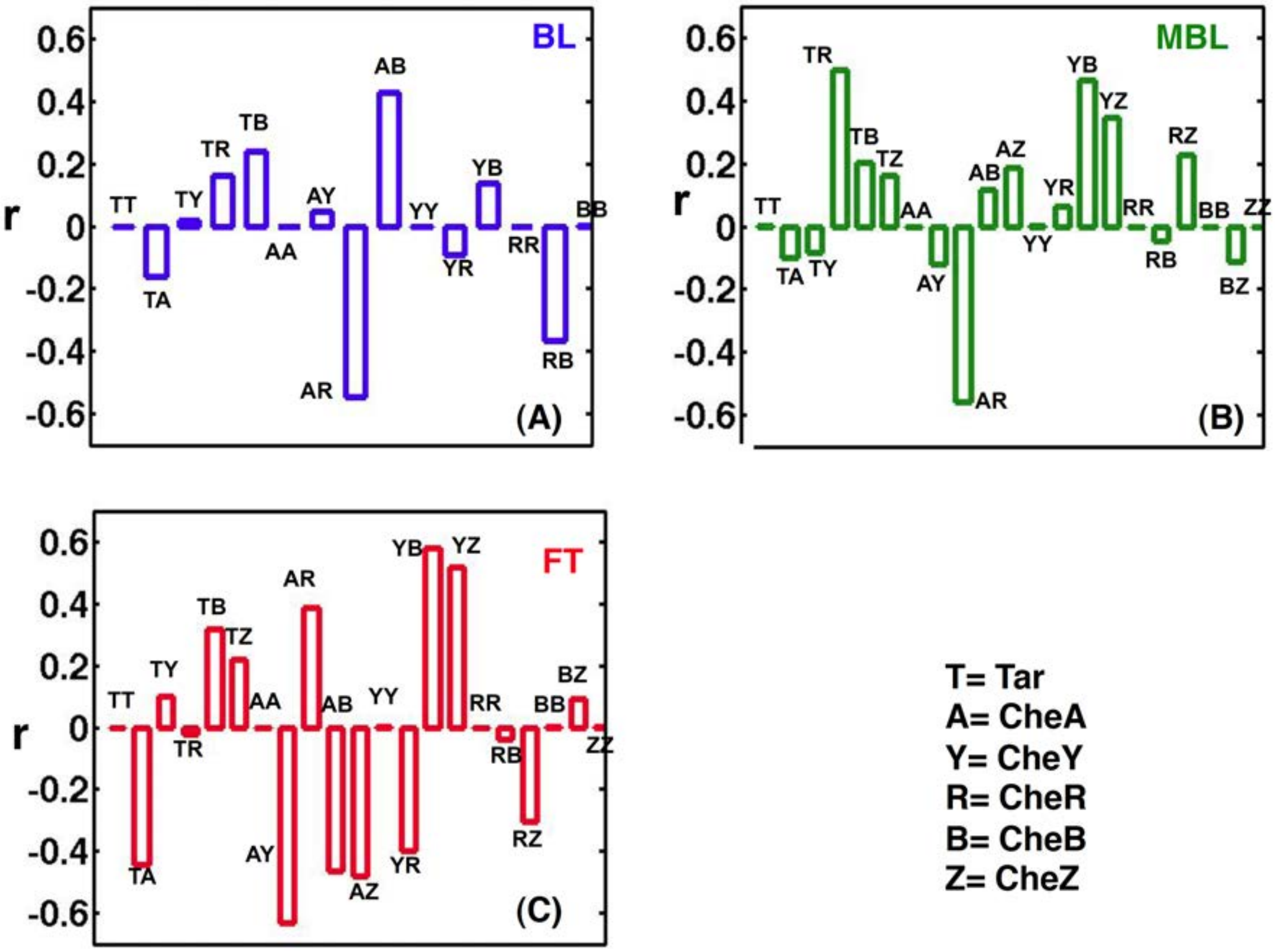

**Fig. S4: Pearson correlations between protein abundances for the BL, MBL and FT models**. We constrained mean values of $\tau$, $\tau^2$, $s$, $p$, and, $p^2$ to the respective values measured in experiments, i.e., $\overline{\tau}$ = 245 s, $\overline{\tau^2}$ = 62323.5 $s^2$, $\overline{s} = 0.02$ , $\overline{p} = 20\%$ , and $p^2$= 425. We show the difference, $r = r_{MaxEnt} - r_{uni}$ for different protein pairs, where, $r_{\mathrm{uni}}$ and $r_{\mathrm{MaxEnt}}$ refer to the Pearson correlation calculated for the *a priori* uniform and the MaxEnt distribution, respectively for **(A)** BL **(B)** MBL **(C)** FT models. The number of samples in all the cases is 70,000.

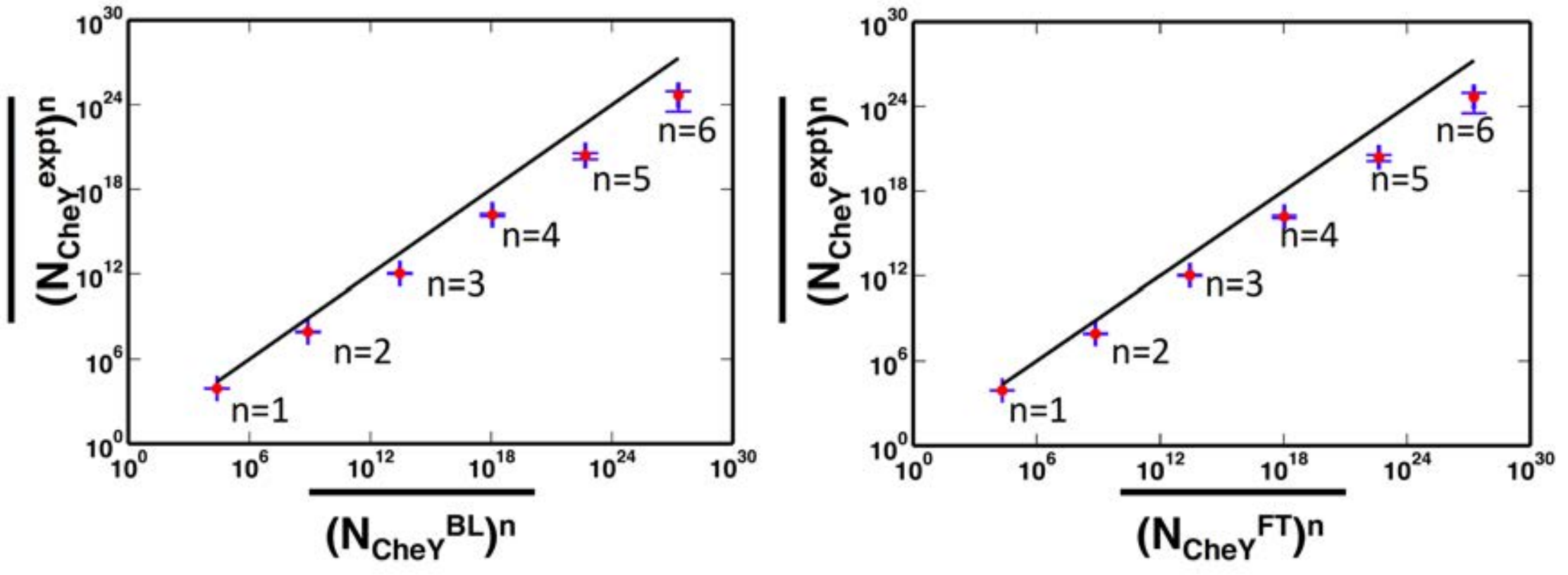

**Fig S5: Comparison between moments of MaxEnt distribution and single cell experiments for the BL (left) and the FT (right) model.** We constrained mean values of τ, $\tau^2$, *s*, *p,* and, $p^2$ to the respective values measured in experiments, i.e., $\overline{\tau}$ = 245 s, $\overline{\tau^2}$ = 62323.5 $s^2$, $\overline{s}$ = 0.02 , $\overline{p}$ = 20% , and $p^2$= 425. The MaxEnt distributions for abundance of CheY for both the models BL and FT, are broader than the experiments. The y=x line (solid black) is shown as a reference. The number of samples used is 70,000.

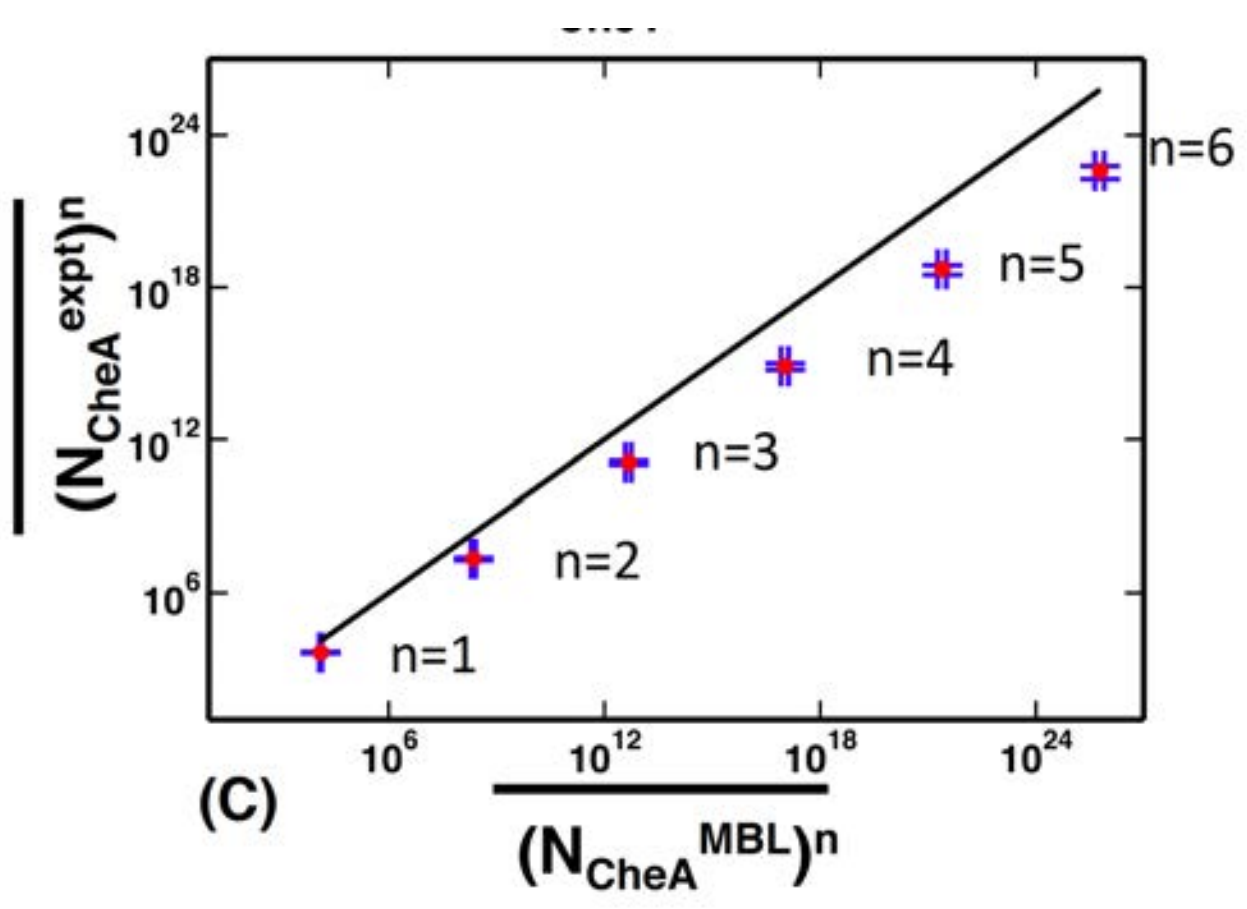

**Fig S6: Comparison between moments of MaxEnt distribution for CheA abundance and single cell experiments for the MBL.** We constrained mean values of τ, $\tau^2$, *s*, *p,* and, $p^2$ to the respective values measured in experiments, i.e., $\overline{\tau}$ = 245 s, $\overline{\tau^2}$ = 62323.5 $s^2$, $\overline{s}$ = 0.02 , $\overline{p}$ = 20% , and $p^2$= 425. The MaxEnt for abundances of CheA for the MBL model is broader than the experiments. The y=x line (solid black) is shown as a reference. The number of samples used is 70,000.

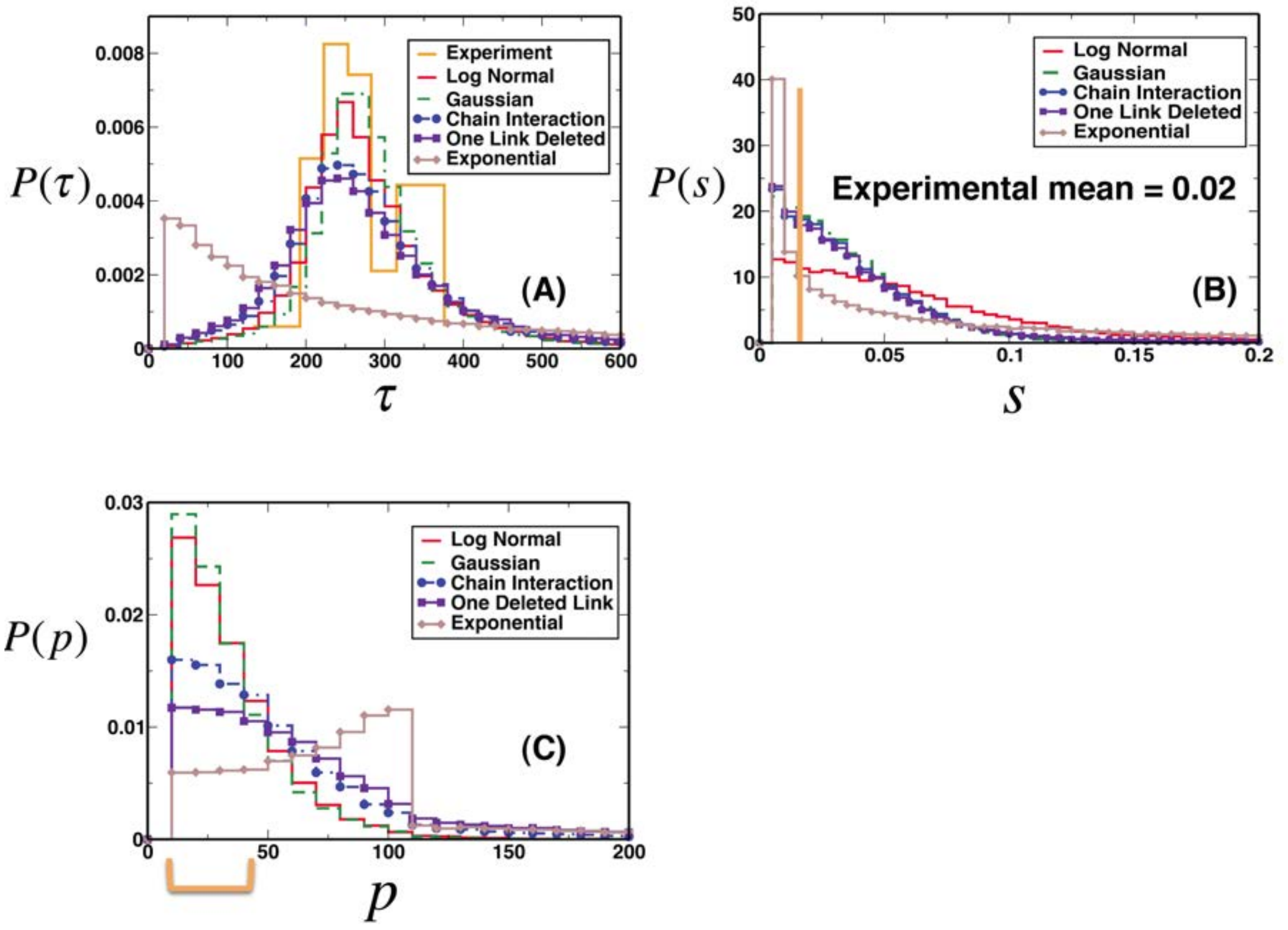

**Fig S7: Distributions of τ, p, and s, when the total protein abundances are drawn from different MaxEnt distributions obtained by imposing different types of constraints on the total protein abundances. (A)** Shows the distribution of the adaptation time when $\{n_q^{total}\}$ are drawn from the Log-normal distribution (**Log Normal**: solid red) as proposed by Kollmann et al. (1); multivariate Gaussian distribution (**Gaussian**: dashed green) where all the pair covariances as well as the mean values are constrained; the Gaussian distribution generated by constraining the covariances between the pairs: (Tar-CheA, CheA-CheY, CheY-CheR, CheR-CheB, and CheB-CheZ), as well as the variances and the mean values of the protein abundances (**Chain Interaction**: dashed blue with solid blue circles); the distribution with the constraints used in the previous case except the pair CheB-CheZ was not constrained (**One Deleted Link**: solid indigo with solid indigo squares); and exponential distribution which was generated when only the mean total abundances were constrained (**Exponential**: solid brown with brown diamonds). The orange graph is the experimental distribution of the adaptation time obtained from Min et al. (25). **(B)** Shows the distribution of the precision of adaptation when the protein numbers for the same types of distributions considered in (A) which are shown with the same color scheme. The orange vertical bar indicates the experimentally measured average precision of adaptation. **(C)** Shows the distribution of the variation of the steady state of the CheY-P protein from an optimal value or *p* for the same types of distributions considered in (A) which are shown with the same color scheme. The orange

square bracket indicates the operational range of the flagella motors. The number of samples used is 37,000 for all the results shown above.

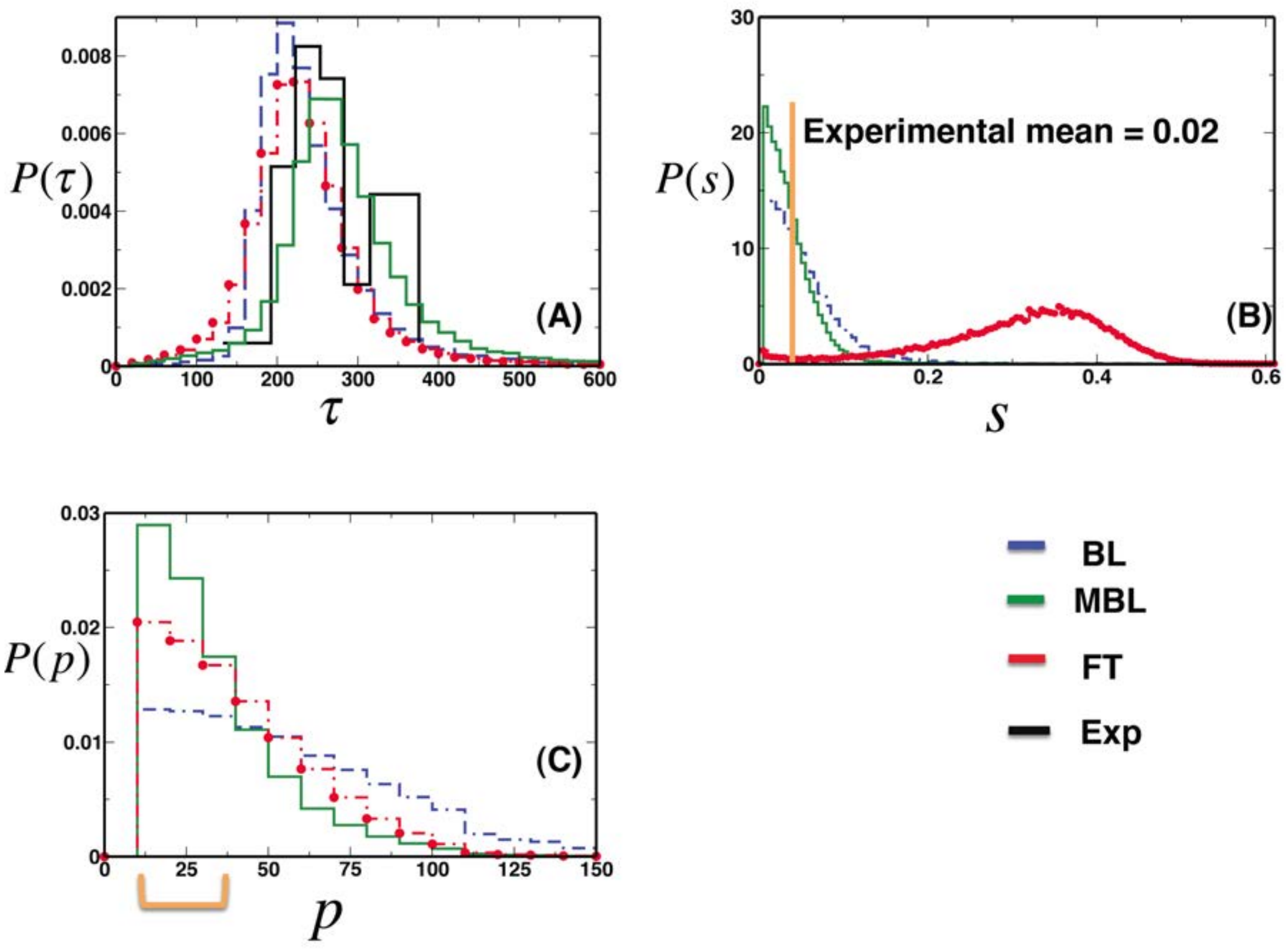

**Fig S8: Shows the distributions of τ, p, and s for BL, MBL and FT models when the mean and the pair correlations of the chemotactic proteins are constrained to the experimental value: (A)** Distribution of the adaptation time for BL (dashed blue), MBL (solid green) and FT model (dashed red with solid red circles) when the mean and the pair correlations between the chemotactic proteins are constrained. The plot in solid black is the experimentally observed adaptation time (25). **(B)** Distribution of the precision of adaptation for BL (dashed blue), MBL (solid green) and FT model (dashed red with solid red circles) when the mean and the pair correlations between the chemotactic proteins are constrained. The vertical line in orange indicates the experimental average precision. **(C)** Distribution of the percentage variation of the steady state of CheY-P from an optimal value for BL (dashed blue), MBL (solid green) and FT model (dashed red with solid red circles) when the mean and the pair correlations between the chemotactic proteins are constrained. The square bracket in orange shows the working range of the flagella motors. The number of samples used is 37,000.

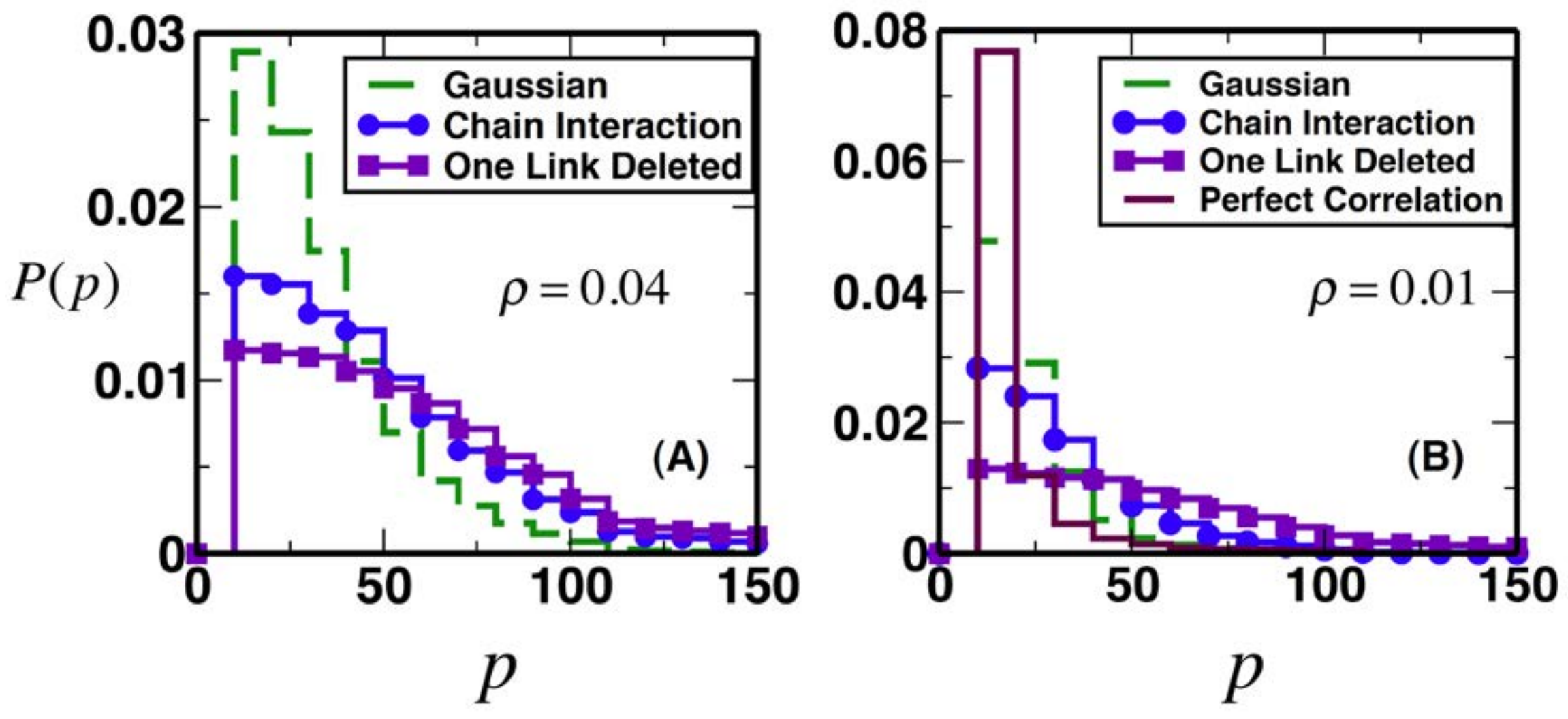

**Fig S9: A comparison of the distribution of *p* for different MaxEnt distributions for two different strength of ρ in Eq. (S16). (A)** Shows the distribution of *p* when the total protein abundances are drawn from a multivariate Gaussian distribution (**Gaussian**: dashed green) where all the pair covariances as well as the mean values are constrained; the Gaussian distribution generated by constraining the covariances between the pairs: (Tar-CheA, CheA-CheY, CheY-CheR, CheR-CheB, and CheB-CheZ), as well as the variances and the mean values of the protein abundances (**Chain Interaction**: dashed blue with solid blue circles); the distribution with the constraints used in the previous case except the pair CheB-CheZ was not constrained (**One Deleted Link**: solid indigo with solid indigo squares) when ρ=0.04 is used in the correlation matrix **C**. **(B)** Shows the same plot as in (A) for ρ=0.01. We also show the case (solid Maroon) when the constrained covariances and variances were taken from the C matrix in Eq. (S17) where ρ=0 (perfect correlation) for reference. The number of samples used is 37,000.

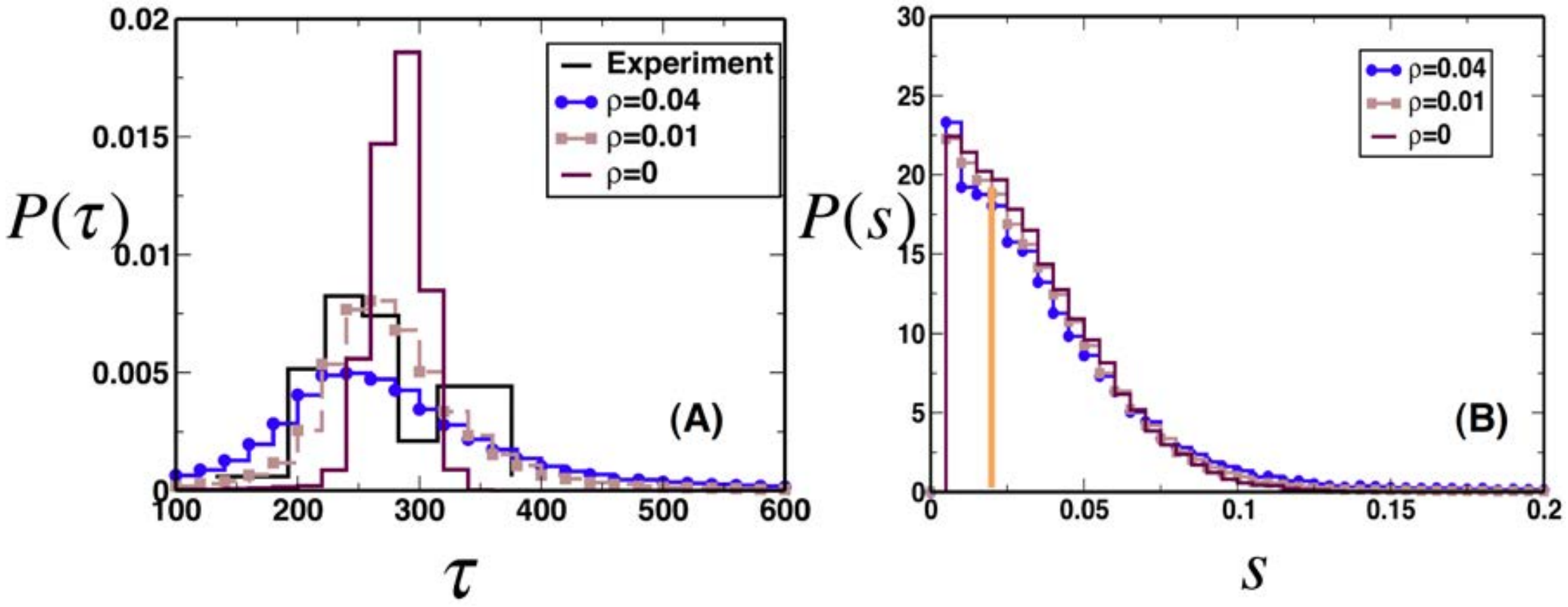

**Fig S10: Comparison of the distribution of the adaptation time and precision of adaptation when the covariances are constrained as shown in Eq. (S16) for different values of ρ. (A)** Adaptation time distribution for the Gaussian distribution generated by constraining the covariances between the pairs: (Tar-CheA, CheA-CheY, CheY-CheR, CheR-CheB, and CheB-CheZ), as well as the variances and the mean values of the protein abundances (**Chain Interaction**) for two different values of ρ=0.04 (blue) and ρ=0.01 (brown). We also show the case (solid Maroon) when the constrained covariances and variances were taken from the C matrix in Eq. (S17) where ρ=0 (perfect correlation) for reference. The experimental distribution is shown in black. **(B)** The distribution of the precision of adaptation for the cases as shown in (A) using the same color scheme. The orange bar indiciates the cell population average value of the precision of adaptation. The number of samples is 37,000.

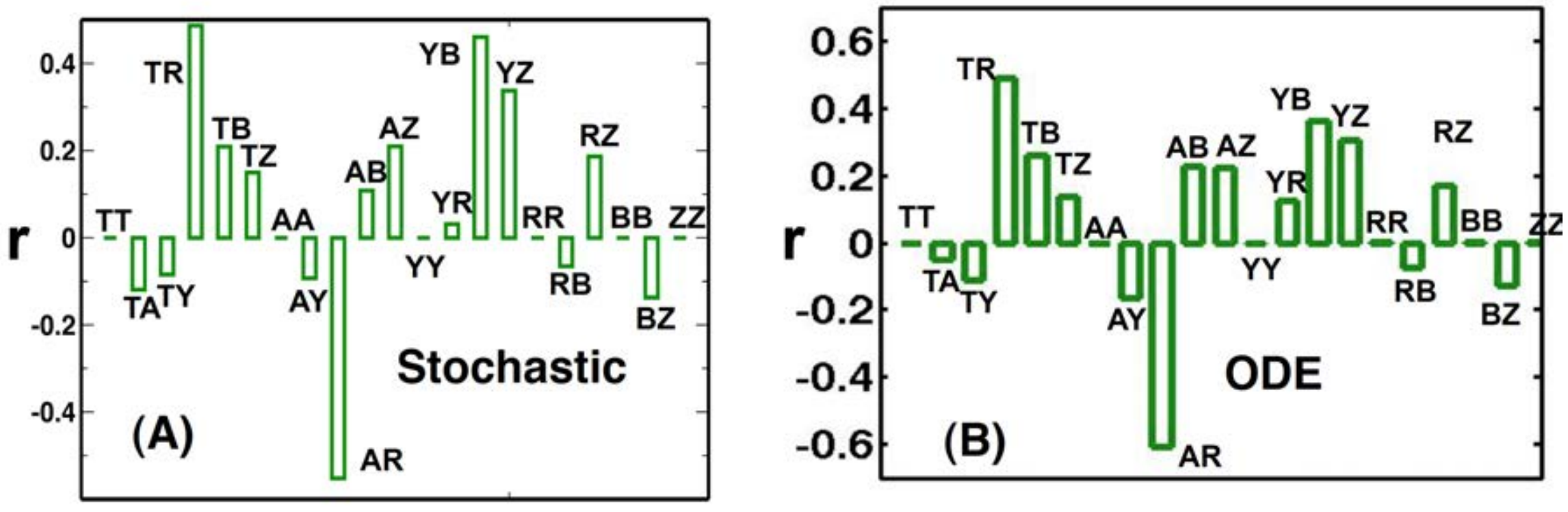

**Fig S11: The effect of intrinsic noise fluctuations on the Pearson correlations between protein pairs.** We constrained mean values of $\tau$, $\tau^2$, $s$, $p$, and, $p^2$ to the respective values measured in experiments, i.e., $\overline{\tau}$ = 245 s, $\overline{\tau^2}$ = 62323.5 $s^2$, $\overline{s} = 0.02$ , $\overline{p} = 20\%$ , and $p^2$= 425 and calculate **(A)** Pearson correlations $r = r_{MaxEnt} - r_{uni}$ among protein pairs using Gillespie algorithm. **(B)** Same as **(A)** but stimulated using ODEs. The number of samples in each case is 70,000. The data is for MBL model.

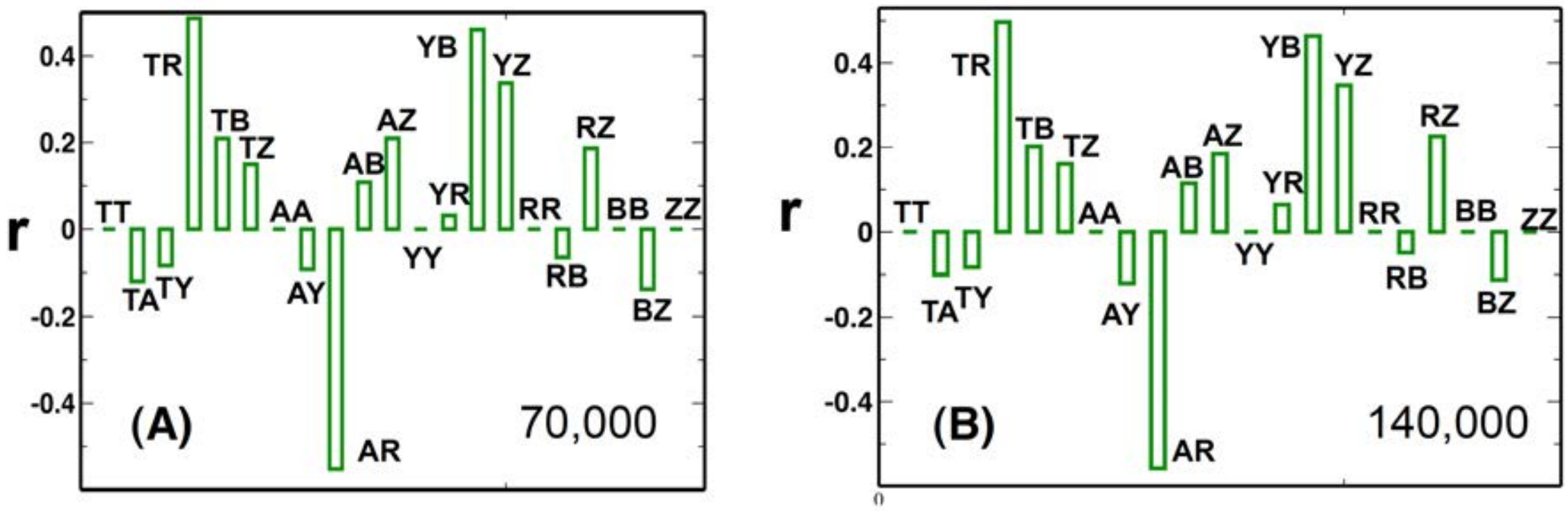

**Fig S12: Convergence of the Pearson correlations for MBL model. (A)** The Pearson correlations are calculated using a sample size of 70,000. **(B)** Same calculation as (A) but done with a sample size of 140,000.

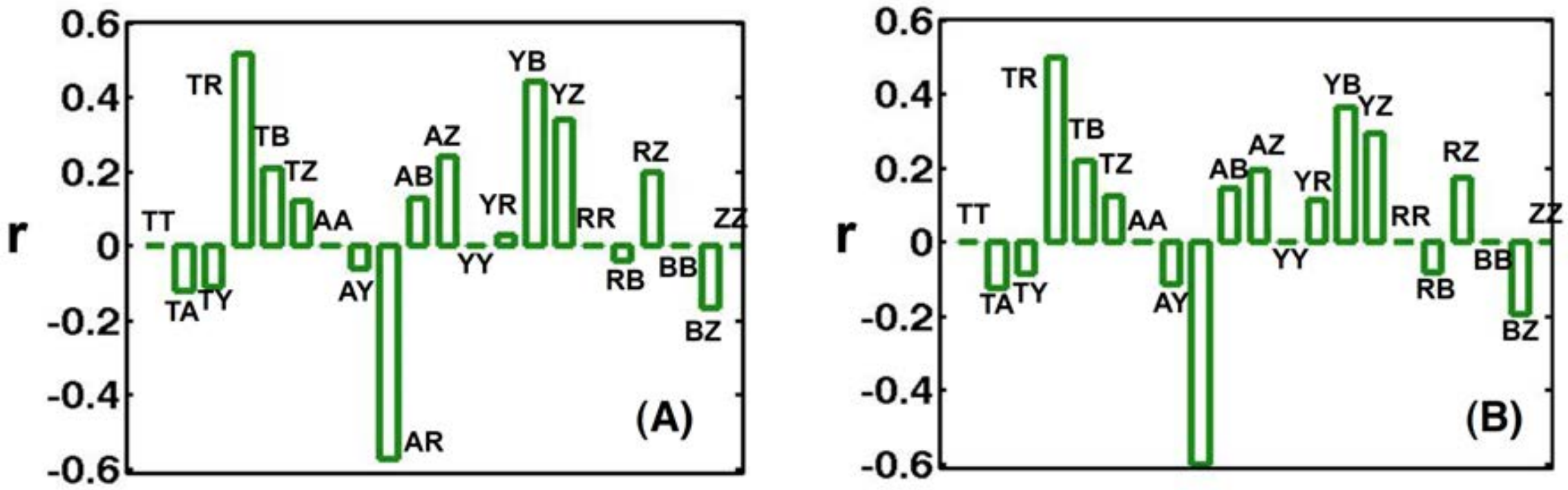

$\bar{\tau} = 245$ s, $\overline{\tau^2} = 62323$ s$^2$, $\bar{p} = 20$ $\overline{p^2} = 425$, $\bar{s} = 0.02$, $\overline{s^2} = 0.0005$

$\bar{\tau} = 263$ s, $\overline{\tau^2} = 71528$ s$^2$, $\bar{p} = 25$ $\overline{p^2} = 725$, $\bar{s} = 0.04$, $\overline{s^2} = 0.0018$

**Fig S13: Sensitivity of the Pearson correlations on the values of the constraints chosen. (A)** The Pearson correlations $r = r_{MaxEnt} - r_{uni}$ are calculated for different protein pairs using the MaxEnt distribution for MBL model. The values of the constraints are shown in the plot. **(B)** Same as (A), but for a different set of values for the constraints. The values used are shown in the plot. The number of samples in each case is 70,000. The data is for MBL model.